\documentclass[manuscript]{aastex}
\usepackage{graphicx}
\usepackage{natbib}
\newcommand{\vlsr}{$V_{\rm LSR}$}
\newcommand{\kms}{~km~s$^{-1}$}

\slugcomment{To be submitted to ApJ}

\shorttitle{Low-Velocity Water Fountains}
\shortauthors{Yung et al.}

\begin{document}


\title{Water Maser Survey on AKARI and IRAS Sources:\\
       A Search for ``Low-Velocity'' Water Fountains}
%\title{Water Maser Survey on Color Selected Evolved Stars:\\ 
%       A Search for ``Low-Velocity'' Water Fountains from\\
%       AKARI and IRAS Sources}


\author{Bosco H. K. Yung\altaffilmark{1}, Jun-ichi Nakashima\altaffilmark{1},
        Hiroshi Imai\altaffilmark{2}, 
        Shuji Deguchi\altaffilmark{3},\\ 
        Christian Henkel\altaffilmark{4}$^{,}$\altaffilmark{5}
        and Sun Kwok\altaffilmark{1}}
%        Philip J. Diamond\altaffilmark{5}$^{,}$\altaffilmark{6}

%\affil{Department of Physics, University of Hong Kong, 
%       Pokfulam Rd, Hong Kong, China}
%\email{byung@hku.hk}

%\author{Hiroshi Imai\altaffilmark{2}, Shuji Deguchi\altaffilmark{3}}
% \affil{Nobeyama Radio Observatory, National Astronomical Observatory,\\
%       and Graduate University for AAppendix materialdvanced Studies, \\
%              Minamimaki, Minamisaku, Nagano 384-1305, Japan}

%\and

%\author{Sun Kwok\altaffilmark{1}}
%\affil{Department of Physics, University of Hong Kong, 
%       Pokfulam Rd, Hong Kong, China}


\altaffiltext{1}{Department of Physics, The University of Hong Kong,
                 Pokfulam Road, Hong Kong, China}
\altaffiltext{2}{Graduate School of Science and Engineering, Kagoshima 
                 University, 1-21-35 Korimoto, Kagoshima 890-0065, Japan}
\altaffiltext{3}{Nobeyama Radio Observatory, National Astronomical Observatory
                 of Japan, Minamimaki, Minamisaku, Nagano 384-1305, Japan}
\altaffiltext{4}{Max-Planck-Institut f{\"u}r Radioastronomie, 
                 Auf dem H{\"u}gel 69, 53121 Bonn, Germany}
\altaffiltext{5}{Astron. Dept., King Abdulaziz University, P.O. Box 80203, 
                 Jeddah, Saudi Arabia}
%\altaffiltext{5}{Jodrell Bank Centre for Astrophysics, Alan Turing Building,
%                 University of Manchester, Manchester M13 9PL, UK}
%\altaffiltext{6}{CSIRO Astronomy and Space Science, PO Box 76, Epping, 
%                 NSW 1710, Australia}


\begin{abstract}
We present the results of a 22~GHz H$_{2}$O maser survey toward a new sample 
of AGB and post-AGB star candidates. Most of the objects are selected for the 
first time based on the AKARI data, which have high flux sensitivity in the
mid-infrared ranges. 
We aim at finding H$_{2}$O maser sources in the 
transient phase between the AGB and post-AGB stage of evolution, where 
the envelopes
start to develop large deviations from spherical symmetry. 
The observations were carried out with
the Effelsberg 100~m radio telescope. Among 204 observed objects, 63 
detections (36 new) were obtained.
We found 4 objects which may be ``water fountain'' sources
(IRAS~15193$+$3132, IRAS~18056$-$1514, OH~16.3$-$3.0, and IRAS~18455$+$0448).
They possess an H$_{2}$O maser velocity coverage much smaller than those in 
other known water fountains. However, the coverage is still larger than that
of the 1612~MHz OH maser. It implies that there is an outflow with 
a higher velocity than the envelope expansion velocity 
(typically $\leq$25\kms), 
meeting the criterion of the water fountain class. 
We suggest that these candidates are possibly oxygen-rich late AGB or early
post-AGB stars in a stage of evolution immediately after the spherically 
symmetric AGB mass-loss has ceased.
\end{abstract}


\keywords{infrared: stars --- masers --- stars: AGB and post-AGB --- 
          stars: evolution --- stars: winds, outflows}



\section{Introduction}
\label{sec:intro}

The circumstellar envelopes of oxygen-rich asymptotic giant branch~(AGB) stars
are well known emitters of OH, H$_{2}$O and SiO masers. 
In particular, 22~GHz H$_{2}$O masers mainly occur in 
the region beyond the dust formation zone. Their spectral profiles are 
sensitive to the mass-loss rate as well as to the geometry of the envelopes 
\citep[e.g.][]{engels86aa,likkel92aa}. Thus, by observing H$_{2}$O masers 
we are able to study in detail the kinematics of the envelopes of evolved 
stars, which are usually optically opaque in short wavelengths.

Maser surveys are commonly performed by single-dish radio telescopes.
However, prior to the launching of the Infrared Astronomical Satellite~(IRAS),
selecting suitable observing targets was not easy due to limited 
methods for identifying evolutionary status. The efficiency has been improved
after \citet{vanderveen88aa} suggested the stellar classification method using 
the IRAS two-color diagram, and since then the IRAS colors have become a main 
criterion used in source selection for many maser surveys. 
The color used is defined as $[m]$$-$$[n]$=$2.5\log(F_{n}/F_{m})$,
where $F_{m}$ and $F_{n}$ represent the fluxes at $m$ and 
$n$~$\mu$m bands, respectively. 
The H$_{2}$O maser survey reported by 
\citet{engels96aas} was a notable early example using the IRAS Point Source 
Catalog. The authors selected over 300 OH/IR stars, which are oxygen-rich 
late AGB stars with very thick envelopes, based on their IRAS [12]$-$[25] and 
[25]$-$[60] colors. They achieved a detection rate up 
to $\sim$50\% using the Effelsberg 100~m radio telescope, and the results
demonstrated the effectiveness of the color selection method. 
Another large-scale H$_{2}$O maser survey was reported by
\citet{lewis97aj}, but here the author focused on color-selected Mira
variables instead of OH/IR stars, and 18 new detections were found from 
$\sim$200 objects. The Haystack 37~m radio telescope was used for this 
observation. A few more H$_{2}$O maser surveys carried out before the year 
2000 are summarized in the Arcetri Catalog 
\citep[][and references therein]{valdettaro01aa}; all of the sources in the 
catalog were observed with the Medicina 32~m radio telescope.
A recent notable survey of similar nature was conducted by
\citet{deacon07apj}. This time the target of interest shifted to the post-AGB
stars. Their searched objects included 85 post-AGB stars
selected by 1612~MHz OH maser properties and IRAS colors.
The Tidbinbilla 70~m radio telescope was used for the maser observation, and 
21 detections were obtained. The most important result from 
\citet{deacon07apj} was the discovery of 3 new ``water fountain~(WF)'' 
sources. There were 9 more post-AGB H$_{2}$O maser sources reported by
\citet{suarez07aa} and \citet{suarez09aa}, which also included 2 new WFs.

WFs are mainly oxygen-rich late AGB or post-AGB stars exhibiting 
high velocity collimated bipolar outflows (or jets) 
traced by H$_{2}$O maser emission \citep[e.g.,][]{imai12pasj}. They are 
suggested to be transitional
objects and their jets contribute to the shaping of planetary nebulae~(PNe) 
\citep[see,][for a review]{imai07iaup,desmurs12iaup}. 
The velocities of their bipolar outflows are larger than those
of the slowly expanding envelopes, which are the remains from the 
mass-loss during the AGB phase. 
In a typical AGB star, the 1612~MHz OH masers show a double-peaked
line-shape. Each peak reveals the velocity of the approaching
blueshifted or the receding redshifted side of the expanding envelope.
OH masers are found beyond the accelerating region, in an environment where 
the expansion motion reaches the terminal velocity
\citep[typically 5--25~km~s$^{-1}$,][]{hekkert89aas}. 
H$_{2}$O masers, on the other hand, mainly show two types of line profile. 
In AGB stars with lower mass-loss (e.g., Mira variables),
the masers show a single emission peak close to the systemic velocity, but in 
the high mass-loss cases (e.g., OH/IR stars), a double-peaked 
profile is observed, similar to that of the OH masers \citep{takaba94pasj}.
In both cases, an H$_{2}$O maser spectral profile would have a smaller 
velocity coverage than that of OH.
For WFs, on the contrary, due to the high velocity (bipolar) outflows,
the spectral velocity coverage of H$_{2}$O is larger than that of OH. WFs are 
rare objects, up to date there are only 15 confirmed members 
\citep{desmurs12iaup}.
Notable examples include IRAS~16342$-$3814 \citep{sahai99apj,claussen09apj},
W~43A \citep{imai02nature} and IRAS~18286$-$0959 \citep{yung11apj}. Most of
the WFs are found to be post-AGB stars, with the 
only possible exceptions of W~43A and OH~12.8$-$0.9 
\citep{boboltz05apj,boboltz07apj}, which may be late AGB stars.

The H$_{2}$O maser spectral velocity coverage of WFs are often very large 
(even $\geq$100\kms). However, a spectrum can only reveal the
line-of-sight velocity of the jet. The velocity coverage is affected by the
inclination angle between the jet axis and the line-of-sight. If the outflow 
direction is nearly perpendicular to it, the maser emission peaks will show
only a small velocity separation in a spectrum, even if the jet has a 
high three-dimensional velocity. 
In fact, OH~12.8$-$0.9 is the WF with the smallest H$_{2}$O velocity 
coverage ($\sim$50\kms), but upon interferometric observations the bipolar
outflow is revealed \citep{boboltz05apj,boboltz07apj}. 
Therefore, some of those ``low velocity'' sources that people have
misidentified as typical AGB stars are probably WFs as well. The smaller 
velocity coverage can be
explained either by a high velocity jet close to the plane of the sky, or 
the objects really possess a slow jet. There have been no studies really 
focusing on these possibilities.

In this paper, we present the results of a new H$_{2}$O maser survey on 
objects selected not only from IRAS, but also from the relatively new AKARI 
database which is characterized by a much higher sensitivity 
(see, Section~\ref{ssec:akari}).
We aim at finding H$_{2}$O masers associated with the aforementioned 
transitional objects and found 4 WF candidates that possess 
even smaller velocity coverage (30--40\kms) than OH~12.8$-$0.9. Their 
expansion velocities are just slightly larger than those for normal AGB stars
($<$30\kms). Thus, without a 
comparison with OH (i.e. the H$_{2}$O velocity coverage should be 
larger than that for OH), it is difficult to notice they could be WFs. 
We suggest the above ``low-velocity'' WF candidates have probably 
reached the late AGB or very early post-AGB phase, which is supported by 
their location in the AKARI two-color diagram (Section~\ref{ssec:nwf}).

The details of the AKARI two-color diagram, sample selection, and the 
observation in this work are given in
Sections~\ref{sec:sam} and \ref{sec:obs}. The results are reported in 
Section~\ref{sec:res}, followed by the discussion in Section~\ref{sec:dis}. 
%The paper is summarized in Section~\ref{sec:con}.

\section{Sample Selection}
\label{sec:sam}

\subsection{AKARI Two-Color Diagram}
\label{ssec:akari}

Most of the observed objects were selected from the AKARI and IRAS databases.
The AKARI data were released in two 
catalogs\footnote{http://www.ir.isas.jaxa.jp/AKARI/Observation/PSC/Public/}. 
The 9 and 18~$\mu$m band fluxes are given in the 
Infrared Camera~(IRC) Point Source Catalog that contains 427,071 objects
\citep{kataza10akari};
the 65, 90, 140, and 160~$\mu$m band fluxes are given in the
Far-Infrared Surveyor~(FIS) Bright Source Catalog, which contains 870,973 
objects \citep{yamamura10akari}. We have used the 9, 18 and
65~$\mu$m band fluxes for object selection because 
these bands have better
sensitivity --- down to 0.045 and 0.06~Jy at 9 and 18~$\mu$m,
respectively 
\citep{kataza10akari}. The band sensitivity at 65~$\mu$m is not given in the 
catalog release note, but that at 90~$\mu$m is given as 0.55~Jy. 
For a comparison, the IRAS sensitivities are about 0.5~Jy at 
12, 25, and 60~$\mu$m, and 1~Jy at 100~$\mu$m.

Using known sources, we define the empirical regions of AGB and post-AGB 
stars on a two-color diagram with AKARI [09]$-$[18] and [18]$-$[65] colors
(Figure~\ref{fig:akari}). The known AGB stars (including Mira variables 
and OH/IR stars) with H$_{2}$O maser detections are selected from 
\citet{engels86aa}, \citet{engels96aas}, \citet{lewis97aj} and 
\citet{valdettaro01aa}. A total of 265 sources are selected. The boundaries 
are set by the outermost AGB stars
distributed on the diagram. For simplicity, a rectangular region is assumed.
The same process is used to find out the color region of
post-AGB stars. We use 38 post-AGB stars with H$_{2}$O maser detections 
to define the color region. In this sample,
30 of them are selected from \citet{deacon07apj}, 
\citet{suarez07aa}, and \citet{suarez09aa}. The remaining 8 are the WFs.
They are selected because most WFs appear to be post-AGB stars 
(see Section~\ref{sec:intro}). Note that only 8 WFs have known AKARI 
fluxes. Therefore, we cannot include all 15 in the AKARI color-color
diagram. Post-AGB stars have a much 
lower H$_{2}$O maser detection rate (see, Section~\ref{ssec:h2o}), and
there were also not so many corresponding observations, hence the smaller 
sample we could obtain.

\subsection{Categories of Observed Objects}
\label{ssec:cate}

A total of 204 observed objects were selected by different 
criteria in addition to the empirical regions of the AKARI two-color diagram 
introduced in Section~\ref{ssec:akari}. In Table~\ref{tab:objs}, the objects 
are put into categories~(a) to (h), according to their nature.
Note that some objects can fall into more than one category.

Category (a). --- Contains potential WFs and known WFs. 
As mentioned in Section~\ref{sec:intro}, the difference between
H$_{2}$O and OH velocity coverages is a key point to distinguish
WFs from other AGB/post-AGB stars. 
Potential WFs are objects, where previous observations
show almost equal velocity coverage of H$_{2}$O and OH masers 
(see the references in Table~\ref{tab:objs}).
These marginal cases might turn out to be WFs once extra H$_{2}$O maser
components are observed at higher velocities. There are 15 objects included.

Categories (b) and (c). --- The former category contains AKARI objects 
(Decl.~$\geq -20^{\circ}$) lying 
inside the post-AGB region in the AKARI two-color diagram 
(Figure~\ref{fig:akari}), 
with $0.5\leq [09]$$-$$[18]\leq 4.5$ and $-0.5\leq [18]$$-$$[65]\leq 2$.
The latter is similar but
for the AGB region in the same diagram, with
$-1\leq [09]$$-$$[18]\leq 1.5$ and $-3\leq [18]$$-$$[65]\leq -0.5$
(Figure~\ref{fig:akari}). There are about 350 and 470 objects fulfilling the 
above criteria for (b) and (c), respectively. However, due to the limited 
observing time, we have selected only those which are relatively bright in
9~$\mu$m ($>$3~Jy). There are finally 68 and 38 objects included.

\begin{figure}[ht]
   \centering
   \includegraphics[scale=0.75]{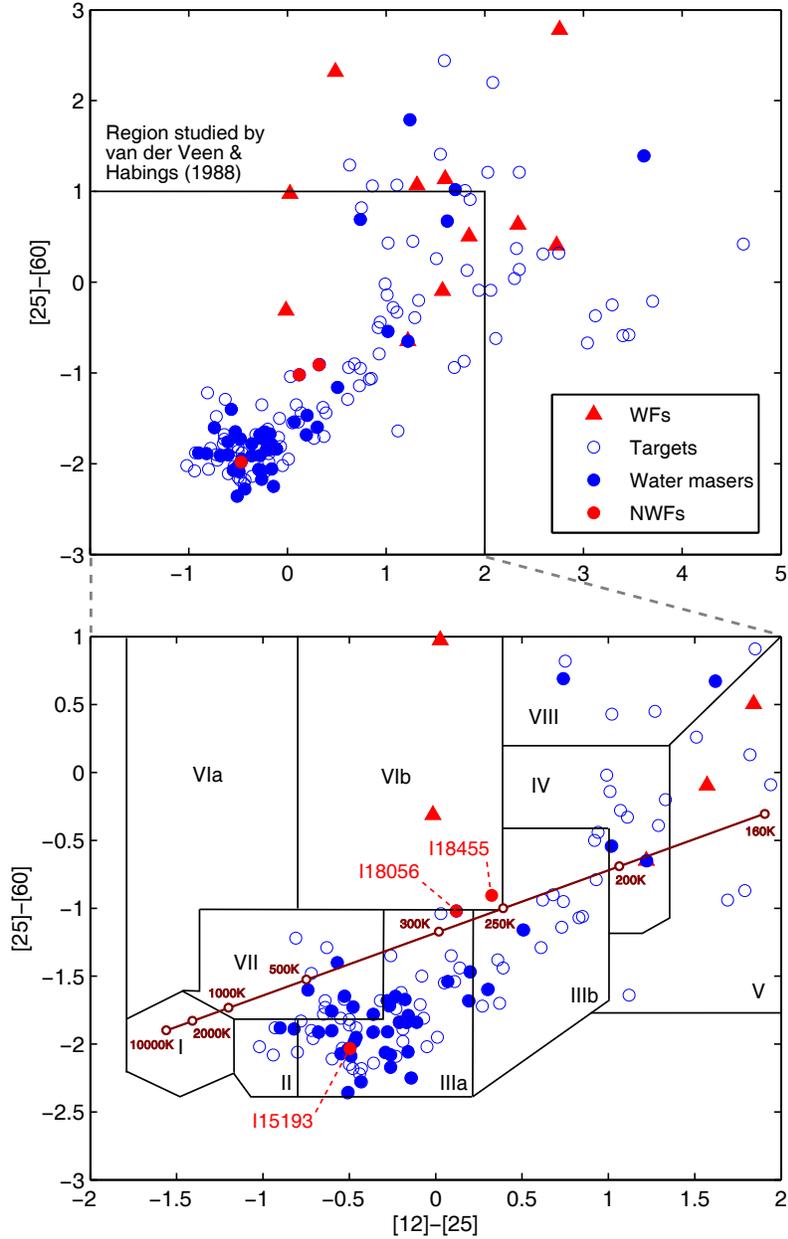}
   \caption{{\bf Upper panel:} IRAS two-color diagram showing 
            the objects observed in the present survey
            (open blue circles), the H$_{2}$O maser detections 
            (filled blue circles),
            the water fountains known to date (WFs; red triangles),
            and 3 of the newly found WF
            candidates (NWFs; filled red circles). We do not
            have sufficient information on the IRAS flux of
            OH~16.3$-$3.0, the forth candidate, thus its position on this
            diagram is not known. {\bf Lower panel:}
            enlarged view of the region studied by 
            \citet{vanderveen88aa}. A blackbody curve is indicated
            by the brown full-line. 
           }
   \label{fig:iras}
\end{figure}

Category (d). --- Contains IRAS objects (Decl.~$\geq -20^{\circ}$) 
selected from the AGB, post-AGB or redder regions (i.e. IIIa, IIIb, IV, V and 
VIII) of the IRAS two-color diagram studied 
by \citet{vanderveen88aa}, as shown in Figure~\ref{fig:iras}. The selection 
method is similar to previous surveys like that of \citet{engels96aas}, but 
new samples are added. There are about 360 objects fulfilling the above
criteria. We have selected only those which are relatively bright 
in 12~$\mu$m ($>$5~Jy). There are finally 106 objects included.  

Category (e). --- Contains known SiO maser sources not observed at 
22~GHz before. SiO masers are often detected in oxygen-rich envelopes. These 
objects are most likely AGB to very early post-AGB stars. There are 98
objects included.

Categories (f) and (g). --- Contains H$_{2}$O non-detections
in two previous major surveys. Category (f) refers to \citet{lewis97aj}, and
(g) to \citet{johnston73apj}. 
As masers show variability, it often happens that previous non-detections 
become new detections upon re-observation some years later. Therefore it is
worth to revisit some of these objects.
It was due to limited observing time that only 6 objects have been observed.

Categories (h). --- Contains other sources that do not belong to any 
of the above categories. These are mainly previously observed objects 
selected from one of the references for re-observation 
(see, Table~\ref{tab:objs}). There are 5 objects included.
%Note: one object can fall into more than one categories!
%(a) potential NWF and known WF: 15
%(b) AKARI WF colors: 68
%(c) AKARI AGB colors: 38
%(d) IRAS AGB, post-AGB, and WF colors: 106
%(e) Deguchi's SiO, not observed or previously non-detected in H2O: 98
%(f) non-detections in Lewis 1997: 1
%(g) non-detections in Johnston et al. 1973: 5
%(h) Other known sources revisited: 5

The columns of Table~\ref{tab:objs} contain the following information:\\
Column 1. --- Object name.\\
Columns 2 and 3. --- R.A. and Decl. in J2000.0.\\
Columns 4 and 5. --- IRAS [12]$-$[25] and [25]$-$[60] colors, respectively.\\
Columns 6 and 7. --- AKARI [09]$-$[18] and [18]$-$[65] colors, respectively.\\
Column 8. --- The local-standard-of-rest velocity~(\vlsr) of the blueshifted 
peak of a double-peaked 1612~MHz OH maser
profile. For a single-peaked profile, the \vlsr\ is recorded in this column
as well, no matter it is really ``blueshifted'' or not.\\
Column 9. --- \vlsr\ of the redshifted peak of a double-peaked 1612~MHz
OH maser profile.\\
Column 10. --- References for the OH maser information given 
in columns 8 and 9.\\
Columns 11 and 12. --- \vlsr\ of the SiO maser peak in the $(v=1$, $J=1-0)$ 
and $(v=2$, $J=1-0)$ transitions, respectively.\\
Column 13. --- References for the SiO maser information given in columns 11 
and 12.\\
Column 14. --- Category, from (a) to (g), to which the object 
belongs.\\

\section{Observation and Data Reduction}
\label{sec:obs}

The observation was performed with the Effelsberg 100~m radio telescope from
2011 November 30 to December 6. An 18--26~GHz HEMT receiver and FFT 
spectrometer were used in the frontend and the backend, respectively. 
The rest frequency of the $6_{16}\rightarrow 5_{23}$ transition line of 
H$_{2}$O molecules was adopted as 22.235080~GHz
\citep{lovas04jphyschem}. 
At this frequency, the full width at half maximum~(FWHM) of the beam was 
about 40\arcsec. 
A 500~MHz bandwidth was used to cover a frequency 
range from 21.985 to 22.485~GHz ($\sim$6700\kms\ at 22.2~GHz). 
The number of spectral channels was 16,384, yielding a channel spacing of
$\sim$0.4\kms. The velocity resolution corresponded to 2 channels,
i.e. $\sim$0.8\kms, which was sufficient to spectroscopically resolve each 
water maser component (normally with a linewidth of $\sim$1--2\kms).
By using the 500~MHz bandwidth, we were able to also detect the H66$\alpha$ 
(with rest frequency 22.364~GHz) and H83$\beta$ (22.196~GHz) recombination 
lines for certain objects (see, Section~\ref{ssec:vel} and 
Appendix~\ref{app:oth}). The velocity scale has been confirmed to be 
accurate by comparisons with previous H$_{2}$O maser spectra of some known 
sources.

An ON/OFF cycle of 2~minutes was used in position switching mode. The OFF 
position was 600\arcsec\ west from the ON position in azimuthal direction. 
The observing time for each source was about 6--20~minutes. The weather was 
fine during most of the observing sessions, 
and the root-mean-square~(rms) noise level was of order 
$10^{-2}$ to $10^{-3}$~Jy.
Pointing was obtained
every 1 to 2 hours with a typical accuracy of about 5\arcsec.
Calibration was obtained from continuum cross scans of sources with flux 
densities given by \citet{ott94aa}.

The data reduction procedures were performed with the Continuum 
and Line Analysis Single-dish Software~(CLASS) package
\footnote{http://www.iram.fr/IRAMFR/GILDAS}. Individual scans on each
object were inspected and those with obvious artifacts were discarded. 
The remaining scans were then averaged. The baseline of each
spectrum was fit by a one-degree polynomial and subtracted, using the channels
without emission.

\section{Results}
\label{sec:res}

\subsection{H$_{2}$O Maser Detections}
\label{ssec:h2o}

In this observation, we obtained a total of 63 detections out of 204 objects, 
and 36 of them are new. Table~\ref{tab:num} gives the number of detections in 
each selection category (see, Section~\ref{ssec:cate} and the last column of
Table~\ref{tab:objs}).
Figure~\ref{fig:new_sp} shows the spectra 
for all the new detections, and Figure~\ref{fig:known_sp} shows the spectra 
for revisited known maser sources. When possible, the velocity range
between the two OH maser peaks and the \vlsr\ of the SiO maser
feature (if present), are also shown. 
The parameters of all the H$_{2}$O maser detections are given in 
Table~\ref{tab:h2o}. The rms noise level of the non-detections are then
given in Table~\ref{tab:non}.

The new attempt of sample selection using AKARI colors yields a detection 
rate of 42\% for the AGB stars (category (c)), but only about 13\% for 
post-AGB stars (category (b)). This drastic drop of the detection rate towards 
redder post-AGB stars is expected, and the same tendency was observed 
in nearly all previous H$_{2}$O maser surveys on evolved stars
\cite[e.g.,][]{engels96aas,valdettaro01aa}.
It is probably due to the decrease in mass-loss rate after the AGB phase,
and by the dissociation of the H$_{2}$O molecules into OH and H
because of the ultra-violet radiation from the central star.
The dissociation also happens at a larger distance from the photosphere,
but this time it is due to external ultra-violet radiation \citep{habing96aar}.
Categories (d) and (e) consist of objects classified by more traditional ways, 
i.e., by their IRAS colors or the SiO maser properties. The detection rate
is about 35\% and 37\%, respectively.

We have 6 additional objects from categories (f) and (g). These objects were
non-detections in previous observations. 
We detected H$_{2}$O masers from 4 of these. 
It indicates that the sample selection criteria used by \citet{lewis97aj} and 
\citet{johnston73apj} were indeed effective.
However, \citet{lewis97aj} achieved only a noise level of $\sim$0.3~Jy using 
the Haystack 37~m radio telescope, and for \citet{johnston73apj}, almost 
40 years ago, the noise level was
up to $\sim$12~Jy for a 5 minute integration using the 26~m reflector of the 
Maryland Point Observatory. 

%Thus sensitivities were 100 to 1000 times worse 
%than that in our current observation, where we have obtained peak fluxes of 
%0.2 to 1~Jy for the objects in question. 
%thus they were not able to detect them mainly due to instrumental
%reasons. Even if we consider masers could have 20--30\% flux variation over 
%time, they are still too weak to be detected under their conditions.

%The parameters of all the detections are given
%in Table~\ref{tab:h2o}. The columns contain the following information:\\
%Column 1. --- Object names.\\
%Columns 2 and 3. --- \vlsr\ and flux density of the blueshifted peak of a
%double-peaked profile. For a single-peaked or irregular              
%profile, the brightest peak is recorded in these two                     
%columns, no matter whether it is really ``blueshifted'' or not.\\
%Columns 4 and 5. --- Same as column 2 and 3, but for the redshifted peak of a
%double-peaked profile, if exist.\\
%Columns 6 and 7. --- \vlsr\ of the two ends of the whole emission        
%profile. The cut-off is defined by the 3-$\sigma$ flux level.\\
%Column 8. --- Integrated flux of the whole emission profile.\\
%Column 9. --- rms noise level.\\
%Column 10. --- References for known detections.

\subsection{Objects with Larger Velocity Coverage of H$_{2}$O than OH Masers}
\label{ssec:vel}

Most of the detections in this project have an H$_{2}$O velocity coverage
smaller than that of OH; this is the behavior of most
circumstellar masers (see, Section~\ref{sec:intro}). 
We found 4 exceptions here. 
These are IRAS~15193$+$3132 (spectrum in Figure~\ref{fig:known_sp}), 
IRAS~18056$-$1514, OH~16.3$-$3.0, and IRAS~18455$+$0448 
(Figure~\ref{fig:new_sp}). 
These objects show an H$_{2}$O velocity coverage which is larger than that
for OH, and therefore they are WF candidates.
Among the 4 objects, IRAS~15193$+$3132 is a known H$_{2}$O maser source,
while the other 3 are new sources.
There is also another object, OH~45.5$+$0.0, with one of its two narrow 
H$_{2}$O features lying outside the OH interval 
(Figure~\ref{fig:new_sp}), but this object is unlikely to be a WF
(to be explained later in this sub-section). 
We have looked at the infrared spectral energy distributions~(SEDs) 
of the 4 candidates. All of them are
characterized by a broad thermal emission feature in the mid- to 
far-infrared range. This is
evidence for the presence of a thick envelope, as radiation coming from the
central star is absorbed and re-emitted at longer wavelengths. A more
extensive study on their SEDs will be presented in 
another paper (Yung et~al. 2013, in preparation). The maser characteristics 
of the above objects are given below.

IRAS~15193$+$3132 (S~CrB). --- 
This is a known H$_{2}$O maser source with a double-peaked profile 
\citep{valdettaro01aa}, which is suggested to be an AGB star with a pulsation 
period of about 360 days \citep{shintani08pasj}. The current observation 
obtained a velocity coverage from about $1$ to $11$\kms. 
This is a very bright source, the S/N is over 1000.
Its OH maser was first detected by \citet{hekkert89aas}, which showed a 
velocity coverage from $-2.5$ to $4.0$\kms. However, in recent OH 
observations obtained in the year 2012 using the Effelsberg 100~m telescope 
(data unpublished), we found that the OH coverage should be $-5.0$ to 
$5.5$\kms\ (Figure~\ref{fig:known_sp}). Nonetheless, it is clear that 
the H$_{2}$O emission exceeds the OH interval on the redshifted side. 
An SiO maser is located at about $-1$\kms\ \citep{kim10apjs}. 
Regarding the distance, the Hipparcos parallex given in the 
Astrometric Catalog (I/311/hip2) is $1.85\pm 1.19$~mas \citep{vanleeuwen07aa}, 
implying a distance of about 540~pc, but with a large uncertainty. 
The total flux calculated from the infrared SED is about
$1.0\times 10^{-10}$~W~m$^{-2}$, corresponding to a luminosity of about 
900~$L_{\odot}$. This value is too low for an AGB star 
(typically $\sim$8000~$L_{\odot}$), and we believe it
is not accurate mainly due to the uncertainty in the parallax distance 
adopted. 
If the distance is confirmed reliable, then the low luminosity could be a 
consequence of a late thermal pulse~(TP), when the star is becoming a 
``born-again AGB'' star \citep[e.g.,][]{blocker95aa}. At the early stage of 
the late TP, the cold envelope is blown away, temporarily exposing the inner 
part of the star. A larger flux from shorter wavelengths is therefore 
expected. Nonetheless, at short wavelengths the influence of interstellar 
extinction is very significant. The total flux observed then decreases, 
and so does the implied luminosity.

IRAS~18056$-$1514. --- 
There are four H$_{2}$O maser peaks associated with
this source, three of them are found within the OH velocity interval
determined by \citet{hekkert91aas}, and one of them is outside. The S/N of
the dimmest peak is about 21 (Figure~\ref{fig:new_sp}).
If we take the central velocity
of the OH masers as the systemic velocity of the star, then the most
blueshifted H$_{2}$O emission (at $\sim$36\kms) implies a projected outflow
velocity of about 23\kms. The expected redshifted peak is missing,
but since masers show variation, the peak is not detected probably 
because it is at its minimum. 
An SiO maser ($v=2$, $J=1-0$ transition only) is detected near the adopted
systemic velocity, at about 
60\kms\ \citep{deguchi00apjs}. The kinematical distance
estimated using the systemic velocity and the galactic rotation curve 
\citep{kothes07aa} is about 4.0~kpc.
Note that the kinematical distance also includes a large uncertainty.
The total infrared flux and the luminosity
are about $4.6\times 10^{-12}$~W~m$^{-2}$ and 2300~$L_{\odot}$, respectively.

OH~16.3$-$3.0. --- 
This object shows a double-peaked profile for both 
H$_{2}$O and OH masers. The S/N of the H$_{2}$O peaks is about 5 
(Figure~\ref{fig:new_sp}). For OH, ``Wing-like'' features 
were found in the outer side of both peaks \citep{sevenster01aa}. 
The velocity coverage between the two OH peaks is about 17\kms, while 
for H$_{2}$O it is about 29\kms. The OH
peaks tell the expansion velocity, so the ``wings'' are probably a sign of
a fast outflow, and the most plausible case is a bipolar outflow.
An SiO maser was found at about $-19$\kms, with the 45~m telescope of the
Nobeyama Radio Observatory~(NRO) on 2012 March 18 (Nakashima 2012,
private communication). It has to be noted that the SiO maser is not located
at the supposed systemic velocity (i.e. the velocity halfway between the
two OH peaks), but it has a similar velocity as one of the OH peaks.
The kinematical distance, total flux and luminosity of the object are
estimated to be 2.6~kpc, $9.0\times 10^{-12}$~W~m$^{-2}$ and 2000~$L_{\odot}$,
respectively.

IRAS~18455$+$0448. --- 
The OH maser profile of this object was first analyzed
by \citet{lewis01apj}. They found that the double-peaked feature was fading 
away over a period of 10 years, and this object has been argued to be a very
young post-AGB star. The OH velocity coverage is about 14\kms. No H$_{2}$O 
masers were detected in the survey conducted by \citet{engels96aas}.
Our new H$_{2}$O spectrum (Figure~\ref{fig:new_sp}) consists of two dominant 
peaks located on each side 
of the OH interval, and one additional blueshifted peak farther away from
the systemic velocity. The S/N of the dimmest peak is about 7, and the total 
H$_{2}$O velocity coverage is about 39\kms.
No SiO masers were detected (Nakashima 2012, private 
communication).
The near and far kinematical distances are 1.9 and 12~kpc, respectively.
\citet{lewis01apj} suggested that in this case the far distance was more 
likely to be correct. 
It is because if one assumes the near distance, the luminosity
would be very low ($\sim$400~$L_{\odot}$), which contradicts to the
post-AGB star status that is supported by other evidence (e.g. by the 
behavior of the 1612 and 1667~MHz OH profiles). Adopting 12~kpc, 
the total flux and luminosity of the object are estimated to be 
$3.6\times 10^{-12}$~W~m$^{-2}$ and 16,000~$L_{\odot}$, respectively.

OH~45.5$+$0.0. --- 
The H$_{2}$O spectrum (Figure~\ref{fig:new_sp}) shows 
two narrow peaks at about 51 and 58\kms, and one of 
them is outside the interval covered by the OH masers (18--53\kms). 
The H66$\alpha$ and H83$\beta$ recombination lines are also detected within
the 500~MHz bandwidth. When the spectra are set to the corresponding rest
frequencies, both lines with broad profiles are centered at about 50\kms\ 
(i.e. agree with the strongest H$_{2}$O feature).

\begin{figure}[ht]
   \centering
   \includegraphics[scale=0.9]{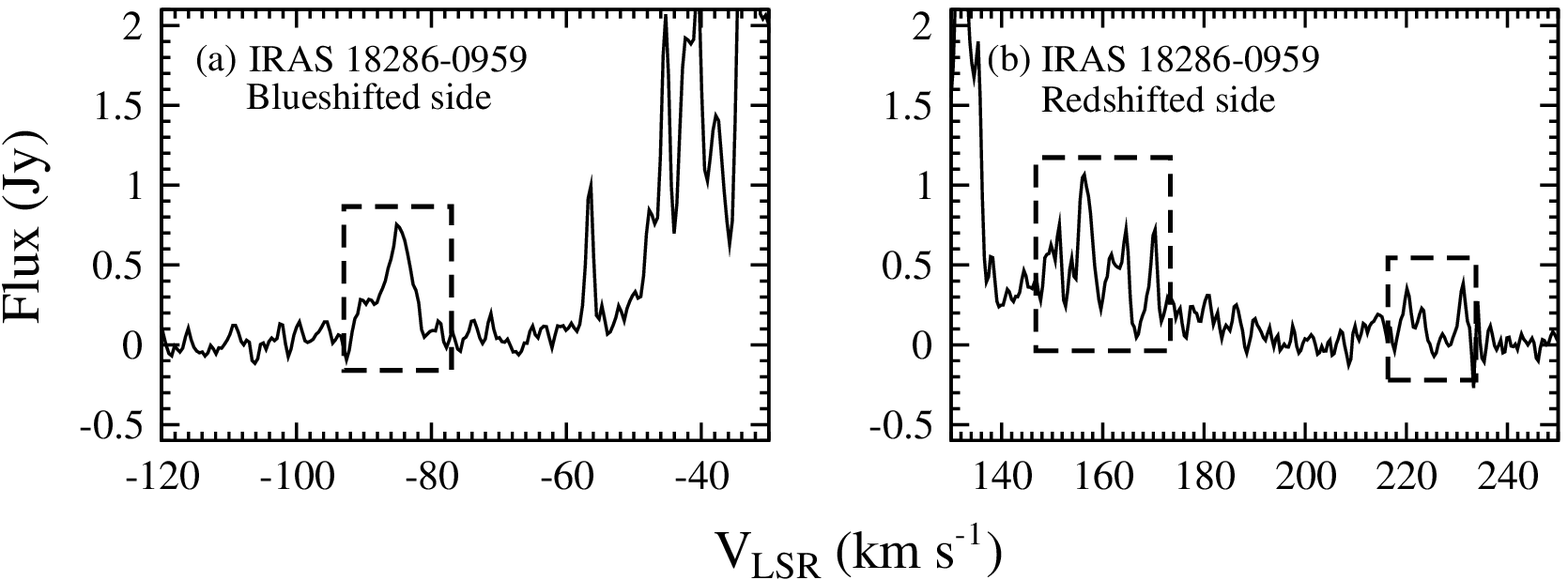}
   \caption{Close-up view of the H$_{2}$O maser spectrum of the WF
            IRAS~18286$-$0959 on the ({\sl a}) blueshifted, and 
            ({\sl b}) redshifted sides. The new velocity components
            are indicated by the broken-line boxes.
           }
\label{fig:i18286}
\end{figure}

In addition, new high velocity H$_{2}$O maser
components are found in the known WF IRAS~18286$-$0959 
(Figure~\ref{fig:known_sp}, and a close up view in Figure~\ref{fig:i18286}), 
when comparing with our previous spectra obtained in 
years 2008 to 2010 with the Very Long Baseline Array~(VLBA) and the 
NRO~45~m telescope.
IRAS~18286$-$0959 has a precessing jet, which exhibits a unique 
``double-helix'' pattern revealed by interferometric observations
\citep{yung11apj}. The H$_{2}$O spectrum has an irregular profile with a lot
of bright peaks.
The new components are weak, but they are still well above the 5-$\sigma$
limit, with a noise level about $5.31\times 10^{-2}$~Jy. The original velocity 
coverage is from about $-60$\kms\ to
$160$\kms. Now we see emission between $-92$\kms\ and $171$\kms. The coverage  
has increased from $220$\kms\ to $263$\kms, 
surpassing IRAS~16342$-$3814 \citep[$\sim$250\kms,][]{claussen09apj}.
The velocity coverage is now the forth largest among the 15 WFs. 
Only the H$_{2}$O emissions from 
IRAS~18113$-$2503 \citep[$\sim$500\kms,][]{gomez11apj}, 
OH~009.1$-$0.4 \citep[$\sim$400\kms,][]{walsh09mnras}, and
IRAS~18460$-$0151 \citep[$\sim$300\kms,][]{deguchi07apj}
are distributed over a wider range.
Therefore, the actual velocity of the jet could be faster than the currently
adopted value \citep[i.e., about 140\kms,][]{yung11apj}, but probably due to 
the variability of the maser flux, these components were not detected before.
Note that the observations committed in 2008 to 2010 have achieved 
a similar sensitivity as the current one (rms~$\sim10^{-2}$~Jy), so that
the possibility of the new peaks being overlooked from previous work is 
excluded.
On the other hand, the jet could indeed accelerate. However, that would
imply an increase of $>$40\kms\ of the jet velocity in 1--2 years. It would
be remarkable if true, because there were no such extreme 
accelerations ever found in evolved stars.
Jet acceleration is possible as it is
already found in the proto-PN CRL~618 \citep{contreras04apj} and another WF
OH~12.9$-$0.9 \citep{boboltz05apj}, but there the acceleration is $<$10\kms\ 
per year. Since the jet of IRAS~18286$-$0959 is precessing, 
it is also possible that the
velocity coverage is affected by the jet direction and shows time variation.
However, according to our kinematical model \citep{yung11apj}, 
an increase of $\sim$40\kms\ is difficult to explain by pure precession
within 2 years.

\subsection{The Unclassified Object 2233550$+$653918}
\label{ssec:unc}

2233550$+$653918 is located in the post-AGB region of the AKARI two-color 
diagram
(Figure~\ref{fig:akari}). It has a near-infrared counterpart in the 2MASS
catalog, but not in IRAS and MSX catalogs. Its WISE image appears to be
a stellar point source. No SIMBAD papers are found regarding this object,
so it is completely new to us. 
It has a double-peaked H$_{2}$O maser profile spreading about 
10\kms\ (Figure~\ref{fig:new_sp}). The S/N are about 16 and 6 for the 
blueshifted and redshifted peak, respectively. The true nature of this object 
is unknown but it might be an evolved star (more in Section~\ref{ssec:evol}).
After the observation, we have found a few more objects with similar infrared
properties (i.e. with post-AGB colors in the AKARI diagram; no IRAS and MSX
counterparts, and no related studies are found). It shows that the high 
sensitivity of AKARI does enable us to find completely new maser sources, 
which could be post-AGB star candidates.

\section{Discussion}
\label{sec:dis}

\subsection{Confirmation of the Evolved Star Status}
\label{ssec:evol}

Some star forming regions~(SFRs) exhibit H$_{2}$O maser
profiles or even infrared colors that resemble those of late-type objects 
such as post-AGB stars, so that misidentifications are possible.
Furthermore, when the objects are close to the Galactic Plane,
contamination by other sources may also occur.
Therefore, first of all, we have to consider the
evolved star status of the newly found WF candidates (especially 
IRAS~18056$-$1514, OH~16.3$-$3.0, and IRAS~18455$+$0448). Their
physical properties as WF are then discussed in later subsections.

We found that there are no other known red sources in IRAS, MSX, AKARI, or 
WISE catalogs except the target sources within the main beam of the telescope,
so the possibility of contamination is firstly excluded. Then, we confirmed 
that there were no 21~cm continuum sources (i.e. H{\sc ii} regions) toward the 
corresponding directions of the 4 WF candidates \citep{condon98aj}. That 
implies these maser sources are not high-mass SFRs. There are also no reports 
on any OH masing low-mass SFR \citep{garay99pasj,sahai07aj}, so the
candidates are unlikely to be low-mass SFRs as well. 
The estimated luminosities given in 
Section~\ref{ssec:vel} might not be accurate due to the 
uncertainties in the adopted distances. However, even in the extreme case of 
having 50\% of distance uncertainty, the luminosities of all those objects are
still brighter than the typical value ($<$100~$L_{\odot}$) for a young stellar 
object.
In addition, they appear as point sources in MSX and WISE mid-infrared images,
which are very different from SFRs that normally show large irregular
extended features. 

Among the other new H$_{2}$O sources, 1817244$-$170623, OH~20.1$-$0.1 
OH~45.5$+$0.0, and OH~70.3$+$1.6 are found
to be lying in H{\sc ii} regions. 
OH~45.5$+$0.0 (Section~\ref{ssec:vel}) and OH~70.3$+$1.6 
(Appendix~\ref{app:oth}) are the only objects toward which the 
H66$\alpha$ and H83$\beta$ recombination lines have been detected. 
These lines are detected in highly ionized regions. 
Note that some recombination lines are also found in PNe 
\citep[e.g.,][]{roelfsema91aa}. 
%For example, H76$\alpha$ and H110$\alpha$ emission lines have been detected
%toward NGC~7027 \citep{roelfsema91aa}. 
Nonetheless, OH~45.5$+$0.0 and
OH~70.3$+$1.6 are unlikely to be PNe because of their highly irregular
shape as shown in mid-infrared images. In addition, PNe are known to have a
very low detection rate of H$_{2}$O masers due to the short lifetime 
($\sim$100 years) of H$_{2}$O molecules in the PN environment 
\citep[e.g.,][]{monsalvo04apj}.
Thus, the above two objects are more likely to be SFRs. 
There are no detailed studies on 1833016$-$105011,
but its mid-infrared images reveal small nebulosity around the central
object. 
%The above sources are possibly contamination to the evolved 
%star samples that could not be removed by color criteria alone.

The rest of the new sources,
including 2233550$+$653918 (Section~\ref{ssec:unc}), are most 
probably evolved objects because of their point-source-like appearance in 
mid-infrared images, 
as well as the non-detection of 21~cm continuum emission. The SiO maser 
detections toward some of the objects provide additional evidence for their 
evolved star status \citep[e.g.,][]{nakashima03pasj2}. The 3 exceptional
cases where SiO masers are detected toward SFRs are 
Orion-KL \citep{kim08pasj}, Sgr~B2 \citep{shiki97apj} and 
W51-IRS2 \citep{morita92pasj}.
Category~(d) of our sample consists of some known SiO maser sources 
(see Table~\ref{tab:objs}).
%SiO masers normally occur in mass-losing AGB stars,
%in the region between the stellar photosphere and the dust formation zone.
%A single-peaked line shape centered close to the systemic velocity is often 
%found. There are only a few exceptional cases where SiO masers are detected 
%toward SFRs, i.e. Orion-KL \citep{kim08pasj}, Sgr~B2 \citep{shiki97apj} and 
%W51-IRS2 \citep{morita92pasj}. Therefore, SiO masers are commonly 
%used to distinguish 
%between evolved stars and SFRs, especially when they have similar infrared 
%colors \citep[e.g.,][]{nakashima03pasj2}. 

\subsection{Properties of the New Water Fountain Candidates}
\label{ssec:nwf}

In Section~\ref{sec:intro}, we mentioned that the smaller H$_{2}$O velocity
coverage of the low-velocity WFs could be just a projection effect on 
high velocity jets, or the jets are intrinsically slower.
For the 4 candidates reported in this paper, 
we suggest below that they are more likely to possess slow jets, and they
are younger than other known WFs in terms of 
evolutionary status. Not knowing the true 
three-dimensional jet velocity yet, we will justify our idea 
by considering infrared colors and maser kinematics. 
Observations using very long baseline interferometry~(VLBI)
will be needed for further analysis.

\subsubsection{AKARI and IRAS Colors}

\begin{figure}[ht]
   \centering
   \includegraphics[scale=0.9]{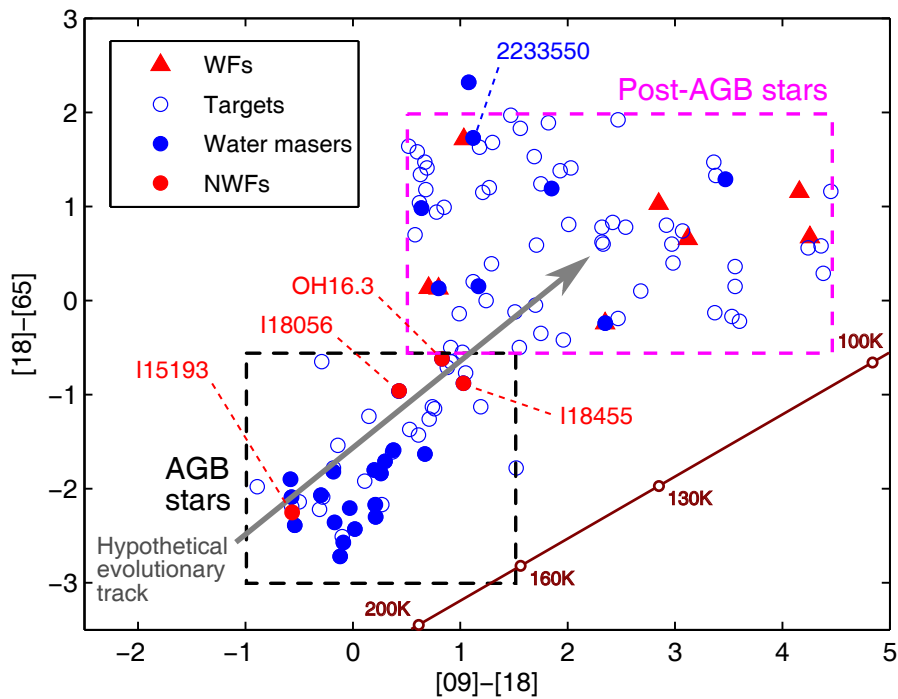}
   \caption{AKARI two-color diagram showing the objects
            observed in the present survey (open blue circles), the
            H$_{2}$O maser detections (filled blue circles), the 
            water fountains known to date (WFs; red triangles), 
            and the newly identified WF candidates 
            (NWFs; filled red circles). The blackbody
            curve is indicated by the brown full-line. The 
            estimated boundaries for AGB and post-AGB stars are
            shown in broken-line boxes. A hypothetical evolutionary 
            track is shown by the grey arrow.}
   \label{fig:akari}
\end{figure}

Based on the distribution of the AGB and post-AGB stars in 
Figure~\ref{fig:akari}, we can assume a rough stellar evolutionary track in
the diagram. 
This is because when an AGB star evolves further, its mass-loss will create a 
very thick dust envelope which obscures the central star. The object will
become very dim, or even unobservable, in optical and near-infrared ranges.
On the contrary, it becomes relatively bright in the mid-infrared due to the 
cold outer envelope. Hence, late AGB and early post-AGB stars normally 
show very red colors
\citep[e.g.,][]{deguchi07apj}. As a result, the evolving AGB stars will move 
toward the ``upper-right'' direction in a two-color diagram, when their
colors become redder.
This evolutionary direction agrees with the model 
prediction by \citet{suh11mnras}.
The same tendency is also noted in the IRAS
two-color diagram presented by \citet{vanderveen88aa}.

From Figure~\ref{fig:akari},
the new WF candidates are expected to be less evolved
than most of the other WFs, but at the same time, at least 3 of them have 
been departed from the main cluster of AGB stars used in our sample.
IRAS~18056$-$1514, OH~16.3$-$0.3 and IRAS~18455$+$0448 were originally
selected by their AKARI colors,
and incidentally, they are located at the upper-right corner of the AGB star
region on the AKARI two-color diagram (see Figure~\ref{fig:akari}), and they 
are not as red as the confirmed WFs in both color indices.
The colors of these 3 candidates indicate that they could be transitional 
objects at the late AGB/early post-AGB stage.
The remaining candidate, IRAS~15193$+$3132, lies in the AGB region of
the AKARI two-color diagram, which is consistent with its suggested AGB 
status.
%Its relatively low luminosity (if reliable) and the hotter color
%could be the result of a late TP as mentioned in Section~\ref{ssec:vel}. 
%If the decrease in flux at longer wavelengths (due to TP), is larger than
%that at shorter wavelengths (due to interstellar extinction), the colors will 
%shift to the bluer side. 

The IRAS two-color diagram suggests a similar story (Figure~\ref{fig:iras}). 
IRAS~15193$+$3132 is found in region IIIa, while IRAS~18056$-$1514 and 
IRAS~18455$+$0448 are found at the boundaries
between region IIIa, IIIb and VIb.
OH~16.3$-$0.3 is missing because the 60~$\mu$m flux is not known.
According to \citet{vanderveen88aa},
variable stars with thick O-rich envelopes are found in region IIIa and IIIb.
These are very likely late AGB stars, where thick envelopes are formed due to 
the mass-loss. Region VIb contains variable stars with hot dust close to
the photosphere, and cold dust at larger distances. These could be
early post-AGB stars, where the steady spherical mass-loss has been 
interrupted, and the dust far away from the central star has cooled down.

\subsubsection{Maser Kinematics}

If we believe that the new candidates are less evolved than other WFs, 
the next question is about their jet velocities. 
The fact that the H$_{2}$O maser velocity coverage is larger than OH implies
a physical differentiation of the faster H$_{2}$O maser flow (probably 
bipolar) from the circumstellar OH flow. From the spectra,
we are not able to determine the three-dimensional velocity. Nonetheless, we
can argue that the chance of the new candidates to be associated with high 
velocity jets is rather low, by considering the orientation of the jet axes. 
Most WFs are found to have a three-dimensional jet velocity in the range of
$100 \leq V \leq 250$\kms\ \citep[e.g.,][]{imai07apj,walsh09mnras,gomez11apj}.
The projected velocity is given by $V\cos(i)$, where $i$ 
(from $0^{\circ}$ to $90^{\circ}$) is
the inclination angle between the jet axis and the line-of-sight. If a
low-velocity WF has an H$_{2}$O maser velocity coverage about 30\kms, then 
the projected velocity of the maser peaks will be 15\kms\ 
(half of the total velocity coverage) off the systemic velocity. 
For jet velocities $V=100$, 150, or 250\kms, we obtain $i\approx 81^{\circ}$, 
$84^{\circ}$, or $87^{\circ}$, respectively. 
Among the candidates that we have found, 3 of them have a velocity coverage 
of even less than 30\kms\ (except IRAS~18455$+$0448). Therefore the 
inclination angle should be at least $81^{\circ}$ for a 100\kms\ jet, 
or even larger for higher velocity jets.

We can estimate the probability of seeing such a jet in the following way.
Assume that there is no bias in the jet axis direction in the three-dimensional
space, so that jets with different orientations are distributed uniformly 
across the sky. Then the probability $P(i_1,i_2)$ of observing a jet with an 
inclination angle between $i_1$ and $i_2$ 
(for $i_1 < i_2$) is given by
\begin{equation}
   P(i_1,i_2) = \frac{1}{2\pi}\int_{0}^{2\pi}
   \int_{i_1}^{i_2}\sin(i)\,{\rm d}i\,{\rm d}\phi~,
\end{equation} 
where $\phi$ is the azimuthal angle defined on the sky plane.
For the extreme projection examples described above 
(i.e. $i_1 = 81^{\circ}, 84^{\circ}$ or $87^{\circ}$; $i_2 = 90^{\circ}$), 
the probabilities calculated by the formula are about 17\%, 10\%, or 5\%. 
On the other hand, the total number of WFs and WF candidates is $15+4=19$. 
Thus, about $5/19=26\%$ of them are low-velocity WFs 
(including OH~12.8$-$0.9), which is higher than the calculated probabilities.
%The chance is even lower for the present case, in which
%4 such objects have been found in a single observation. 
Hence, the small H$_{2}$O velocity coverages are probably not caused by
pure geometrical effects, and the objects are likely to have
intrinsic slow jets.

According to the known cases of jet acceleration in 
OH~12.8$-$0.9 and CRL~618 (Section~\ref{ssec:vel}), we could assume that 
the very ``first'' bipolar outflow from a late AGB star might actually 
occur with a lower velocity, then it gradually accelerates. 
This is consistent with our interpretation that the new WF candidates are
younger and possess slower jets than other known WFs.
The acceleration mechanism is still not clear, but there exist mechanisms 
like that proposed by the magnetocentrifugal launching ~(MCL) model
\citep[see,][and references therein]{dennis08apj}.
The MCL model assumes the system has a rotating central gravitating object
(the central star),
which may or may not include an accretion disk. For the case that the disk
is present (which is quite common in post-AGB stars), 
plasma is threaded by a magnetic field whose poloidal component is 
rotating in the same direction as the disk. The ionized gas of the disk is 
then subject to centrifugal force and accelerates. The gas is thrown out 
along the field lines. As the system expands, the toroidal component of the 
field dominates and the hoop stress (or circumferential stress) 
collimates the outflow. 
Note that a magnetic field ($B\approx 200$~mG at one position along the jet) 
has already been detected toward W~43A, the first 
candidate of the WF class \citep{vlemmings06nature}. Therefore, it is possible 
that magnetic fields play a major role in collimating as well as 
accelerating the jets. 
If such acceleration is occurring,  
then the existence of the low-velocity WF candidates will imply that the 
dynamical age of WFs should be much larger than what has been expected 
\citep[less than 100 years,][]{imai07iaup}, because the current adopted value 
was estimated only with the high jet velocities.

Finally, regarding the SiO masers of the new WF candidates, we note that
IRAS~18056$-$1514 has an SiO emission line at the systemic velocity.
The fact that only the $v=2$ line has been detected is consistent with
the late AGB/early post-AGB phase prediction, as \citet{nakashima03pasj2} 
found that the SiO $v=2$ line will become dominant as the objects get redder 
in their IRAS colors (i.e. more evolved objects). 
IRAS~15193$+$3132 also exhibits an SiO maser feature at about the systemic
velocity, but only the $v=1$ line has been measured (Table~\ref{tab:objs}).
The behavior of its SiO and OH masers also agree with its AGB status.
It implies that the onset of an asymmetric outflow (as indicated by the 
H$_{2}$O and OH maser profiles) could actually happen at a stage much earlier 
than the post-AGB phase. The other 2 candidates show either SiO 
non-detections, or
the emission peak is seen off the systemic velocity. We suggest this is 
due to the preliminary morphological change, when the envelopes start to 
develop bipolarity. The SiO masers at this stage are probably originated from 
an elongated region, and a double-peaked profile is expected even though we 
only have a single peak
for OH~16.3$-$3.0. A similar example is the 
bipolar SiO outflow of the late AGB star W~43A, which has been mapped with 
the Very Large Array \citep{imai05apj}.
This stage, however, is expected to be short. As the star evolves
further and the envelope is detached from the star the SiO maser will 
disappear. This could be the case of IRAS~18455$+$0448
and the rest of the WFs. 
Therefore, considering all the properties discussed, the new WF candidates 
could be characteristic representatives of the short
transition stage at the late AGB/early post-AGB phase, when the morphology 
of the envelopes starts to develop asymmetry.

\section{Conclusions}
\label{sec:con}

We have conducted a 22~GHz water maser survey on 204 objects, mainly AGB
and post-AGB stars, using various source selection criteria such as the
AKARI two-color diagram. There are 63 detections and 36 of them are
new, including an unclassified object that was first identified by the 
AKARI observations, 2233550$+$653918. 
New high velocity components are also found in the known ``water
fountain'' IRAS~18286$-$0959.
We have found 4 new candidates for this water fountain class, but having
much smaller H$_{2}$O maser velocity coverage than other known examples.
In principle, the smaller velocity coverage could just be a projection effect, 
or the objects are really associated with slower jets. 
From our statistics, we suggest that they are more likely to have intrinsic
slow jets. They could be transitional objects undergoing a 
morphological change, during the late AGB/early post-AGB stage. 
Studying the kinematical process occurring at this stage is helpful for us to 
understand the shaping of planetary nebulae. 
Nonetheless, the true status of the candidates can only be confirmed upon 
interferometric observations, or by high resolution infrared imaging, to see 
whether there are bipolar structures or not. The three-dimensional velocity
of the outflow could also be determined by measuring the proper motions of the
maser features, with multi-epoch VLBI observations 
\citep[e.g.,][]{imai02nature,yung11apj}.


%% Included in this acknowledgments section are examples of the
%% AASTeX hypertext markup commands. Use \url without the optional [HREF]
%% argument when you want to print the url directly in the text. Otherwise,
%% use either \url or \anchor, with the HREF as the first argument and the
%% text to be printed in the second.

\acknowledgments

%This research is supported by grants from the Research Grants Council
%of Hong Kong
%(project code: HKU~703308P, HKU~704209P, HKU~704710P, and HKU\,704411P),
%and 
%the Small Project Funding of The University of Hong Kong
%(project code: 201007176004). 
%B.Y. acknowledges the support by the HKU SPACE Research Fund. 
%H.I. is financially supported by Grant-in-Aid for Young Scientists from 
%the Ministry of Education, Culture, Sports, Science, and Technology 
%(18740109), as well as by Grant-in-Aid for Scientific Research from Japan 
%Society for Promotion Science (20540234).
%The National Radio Astronomy Observatory is a facility of the National Science
%Foundation operated under cooperative agreement by Associated
%Universities, Inc.
This work is supported by a grant awarded to J.N. from the Research
Grants Council of Hong Kong (project code: HKU 704710P) and the Small
Project Funding of the University of Hong Kong (project code:
201007176004).
The results are based on observations with the 100~m telescope of the 
MPIfR (Max-Planck-Institut f{\"u}r Radioastronomie) at Effelsberg, and AKARI, 
a JAXA project with the participation of ESA.


%% in parentheses, after the keyword, but they will not be verified.

%{\it Facilities:} \facility{Nickel}, \facility{HST (STIS)}, \facility{CXO (ASIS)}.



\appendix

\section{Other Notable H$_{2}$O Maser Detections}
\label{app:oth}

There are some notable sources in addition to the new water fountain~(WF) 
candidates found in this project, which will be briefly described below: 

IRAS~06319$+$0415 (RAFGL~961). --- This is a well known object suggested to
be a massive protostar, made famous by the detection of the
water ice vibrational band \citep[e.g.,][]{smith11mnras}. It is however, the
first time that an H$_{2}$O maser is found.

OH~70.3$+$1.6. --- It has two H$_{2}$O maser peaks. 
We suggest it is a high mass star forming region~(SFR), 
as the H66$\alpha$ (22.364~GHz) and 
H83$\beta$ (22.196~GHz) recombination lines are detected within our 500~MHz 
bandwidth. Both lines are centered at about $-30$\kms\ (i.e. roughly halfway
between the velocities of the two H$_{2}$O peaks). These lines only occur in 
highly ionized H{\sc ii} region.
The object also shows characteristic extended features in mid-infrared images,
which agrees with the SFR assumption.

IRAS~19271$+$1354. --- This object has two clusters of H$_{2}$O peaks. The
detection of the blueshifted cluster is reported in \citet{engels96aas}, and 
the present observation is the first time that the more redshifted cluster is 
detected. 
A single-peaked feature was found in both the OH \citep{chengalur93apjs} 
and SiO \citep{nakashima03pasj1} spectra, and the velocities of both emission 
peaks are lying outside the H$_{2}$O velocity range. The SiO maser 
resembles that of the new WF candidate OH~16.3$-$3.0
(i.e. significantly drifted away from
the assumed systemic velocity). The object might have started to
develop asymmetry in the very inner part of the envelope. 
Nonetheless, lacking fully convincing evidence, we conservatively
do not suggest it is a WF candidate. To prove its true status, we have to at
least detect both OH peaks and get the accurate envelope expansion velocity.

IRAS~19295$+$2228. ---
This object is identified as an OH/IR star, and it is visually close 
($\sim$130\arcsec) to another object with similar nature, IRAS~19296$+$2227. 
Their OH masers were
observed in the same beam and recorded with the designation OH~57.5$+$1.8.
However, \citet{engels96aa} found that the 2 clusters of OH masers, with 
very different line-of-sight velocities, actually belong to 2 different 
sources. H$_{2}$O maser emission was found in IRAS~19296$+$2227, but not in 
IRAS~19295$+$2228 \citep{engels96aas}. Therefore, here we present the first 
detection in H$_{2}$O toward IRAS~19295$+$2228. In addition, 
\citet{nakashima03pasj2} searched for 43~GHz SiO masers toward both objects, 
but it was only found in IRAS~19295$+$2228.

IRAS~19312$+$1950. --- A new H$_{2}$O peak at 26\kms\ is added to the known 
double-peaked profile of this source. 
Currently there are several speculations about the true nature of this 
peculiar object:
it could be a post-AGB star embedded in a small dark cloud by chance,
a red nova formed by a merger of two main sequence stars, or a coincidence
of a background/foreground small dark cloud appearing in the direction
of the IRAS source with the same \vlsr. Upon interferometric observations, 
it is shown that the two original H$_{2}$O 
peaks correspond to a possible bipolar outflow 
\citep[see,][for a detailed study of this object]{nakashima11apj}. 
It is unclear how the new peak is produced in the system.

\bibliography{ms}

%% the file, uncomment the \plotone, \plottwo, and
%% If you need a layout that cannot be achieved with \plotone or
%% \plottwo, you can invoke the graphicx package directly with the
%% \includegraphics command or use \plotfiddle. For more information,
%% please see the tutorial on "Using Electronic Art with AASTeX" in the
%% documentation section at the AASTeX Web site,
%% http://www.journals.uchicago.edu/AAS/AASTeX.
%%
%% The examples below also include sample markup for submission of
%% supplemental electronic materials. As always, be sure to check
%% the instructions to authors for the journal you are submitting to
%% for specific submissions guidelines as they vary from

%% This example uses \plotone to include an EPS file scaled to
%% 80% of its natural size with \epsscale. Its caption
%% has been written to indicate that additional figure parts will be
%% available in the electronic journal.

%% Here we use \plottwo to present two versions of the same figure,
%% one in black and white for print the other in RGB color
%% for online presentation. Note that the caption indicates
%% that a color version of the figure will be available online.
%%

%\begin{figure}
%\plottwo{f2.eps}{f2_color.eps}
%\caption{A panel taken from Figure 2 of \citet{rudnick03}. 
%See the electronic edition of the Journal for a color version 
%of this figure.\label{fig2}}
%\end{figure}

%%%%%%%% My figures %%%%%%%%%%%%%

\clearpage

%%%%%%%%%%%%%%%%%%%%%%% Figure 1 %%%%%%%%%%%%%%%%%%%%%%%%%%%

%%%%%%%%%%%%%%%%%%%% Figure 1 Ends %%%%%%%%%%%%%%%%%%%%%%%%%

%%%%%%%%%%%%%%%%%%%%%%% Figure 2 %%%%%%%%%%%%%%%%%%%%%%%%%%%

%%%%%%%%%%%%%%%%%%%% Figure 2 Ends %%%%%%%%%%%%%%%%%%%%%%%%%

%%%%%%%%%%%%%%%%%%%%%%% Figure 3 %%%%%%%%%%%%%%%%%%%%%%%%%%%
\begin{figure}[ht]
   \centering
   \includegraphics[scale=0.9]{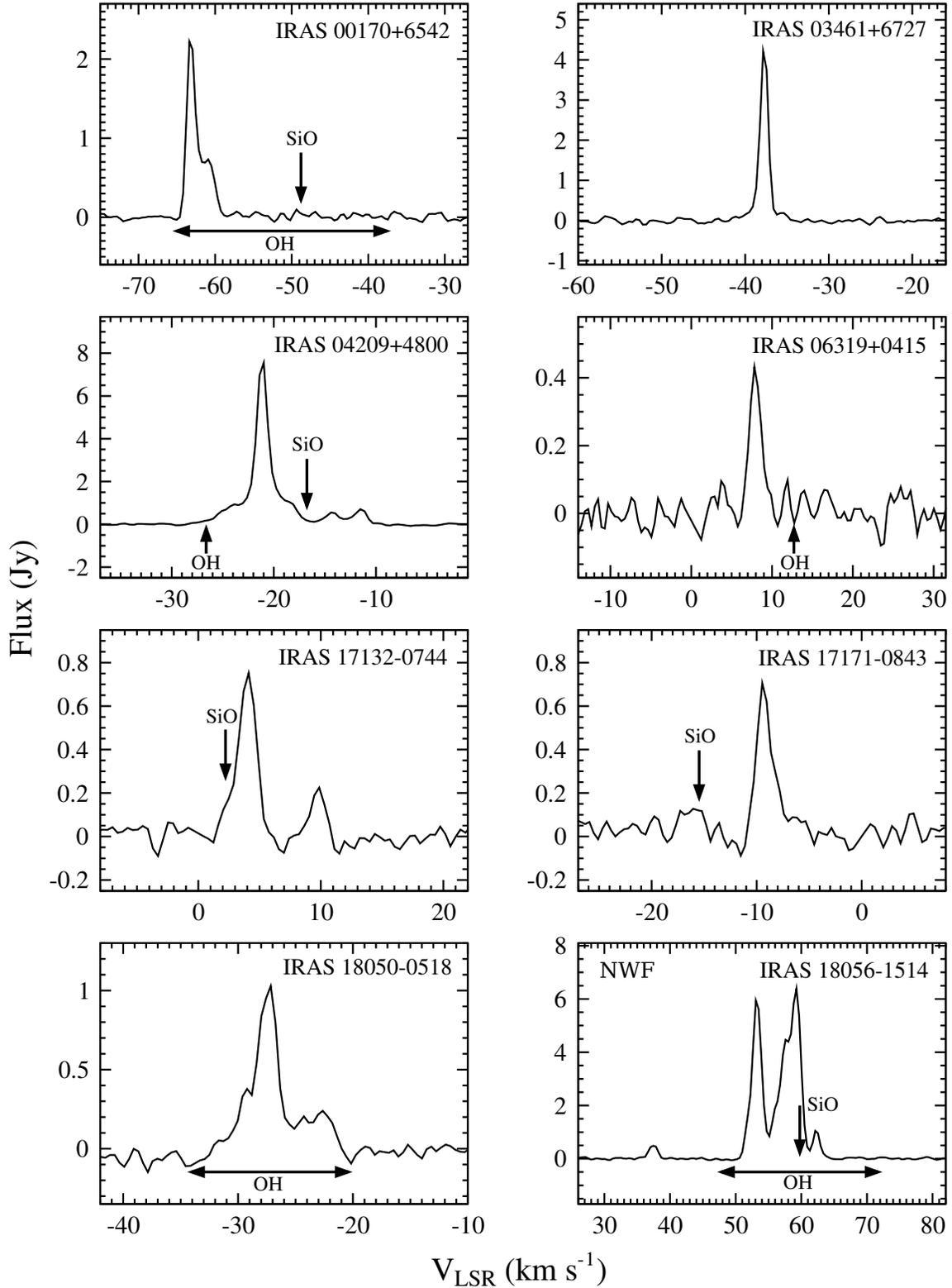}
   \caption{Spectra of new H$_{2}$O maser detections. 
            The new ``low-velocity'' water fountain candidates
            (IRAS~18056$-$1514, OH~16.3$-$3.0 and IRAS~18455$+$0448) 
            are labeled as ``NWF''.}
   \label{fig:new_sp}
\end{figure}

\begin{figure}
   \figurenum{3}
   \centering
   \includegraphics[scale=0.9]{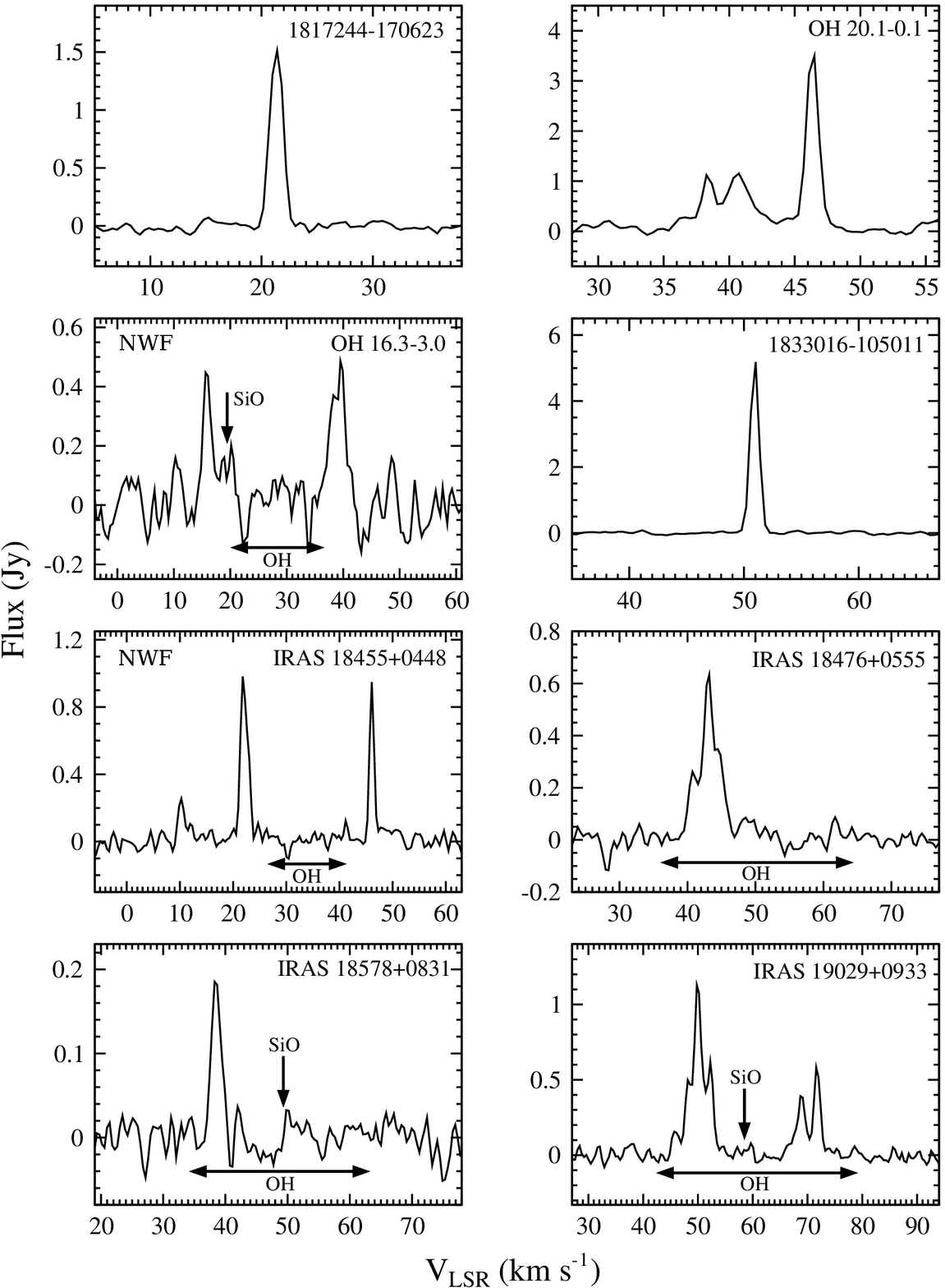}
   \caption{\it Continued}
\end{figure}

\begin{figure}
   \figurenum{3}
   \centering
   \includegraphics[scale=0.9]{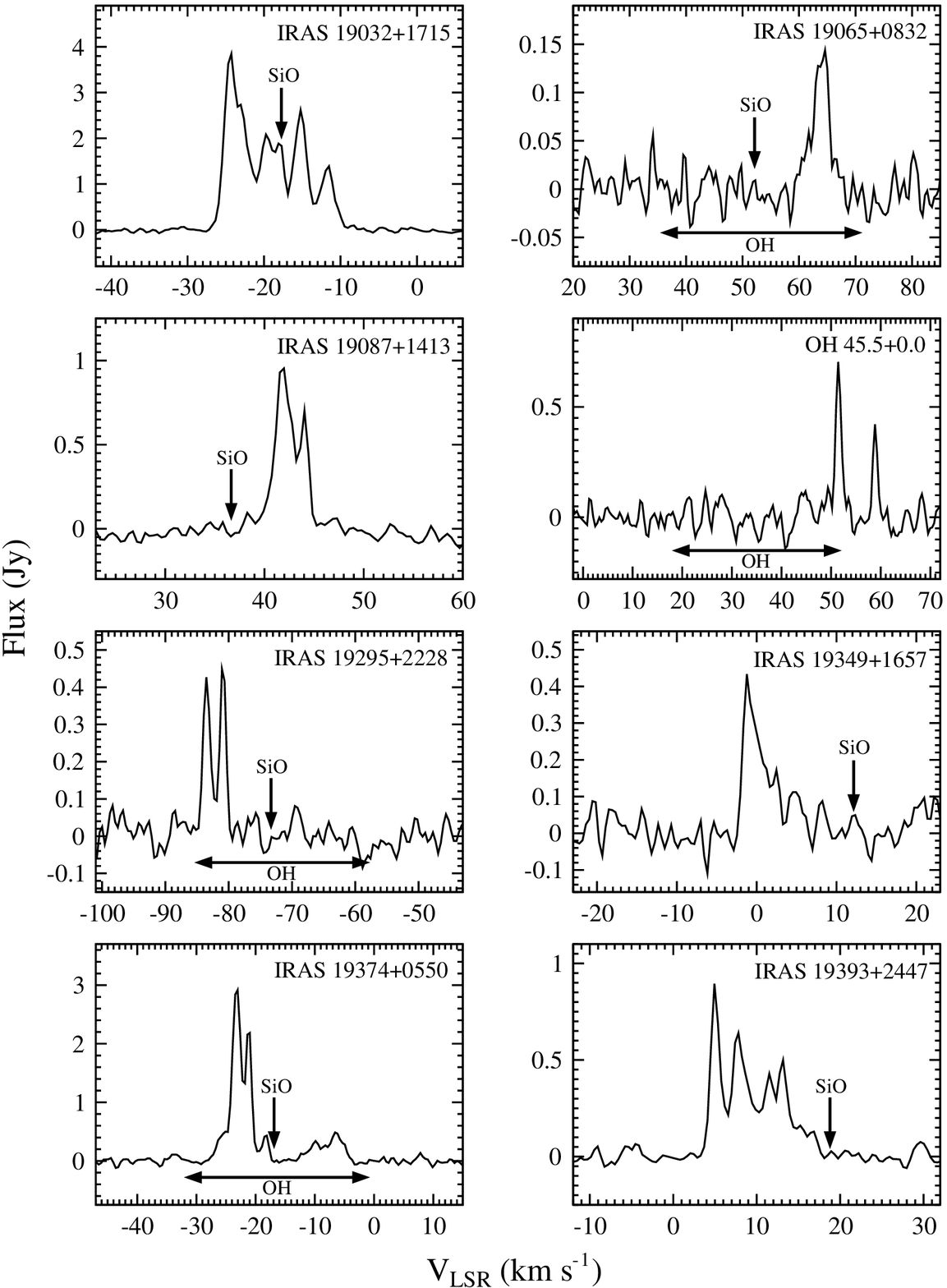}
   \caption{\it Continued}
\end{figure}

\begin{figure}
   \figurenum{3}
   \centering
   \includegraphics[scale=0.9]{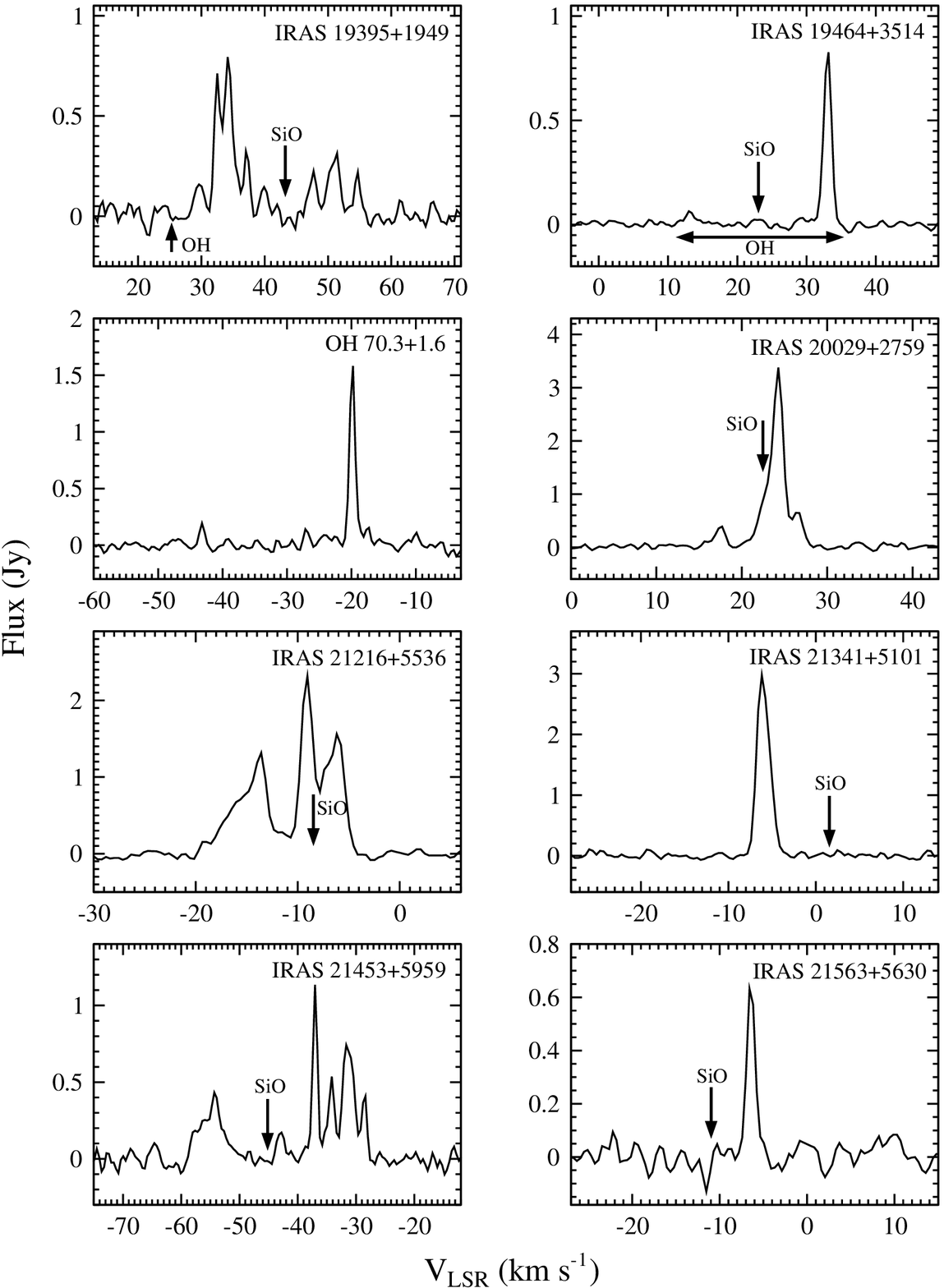}
   \caption{\it Continued}
\end{figure}

\begin{figure}
   \figurenum{3}
   \centering
   \includegraphics[scale=0.9]{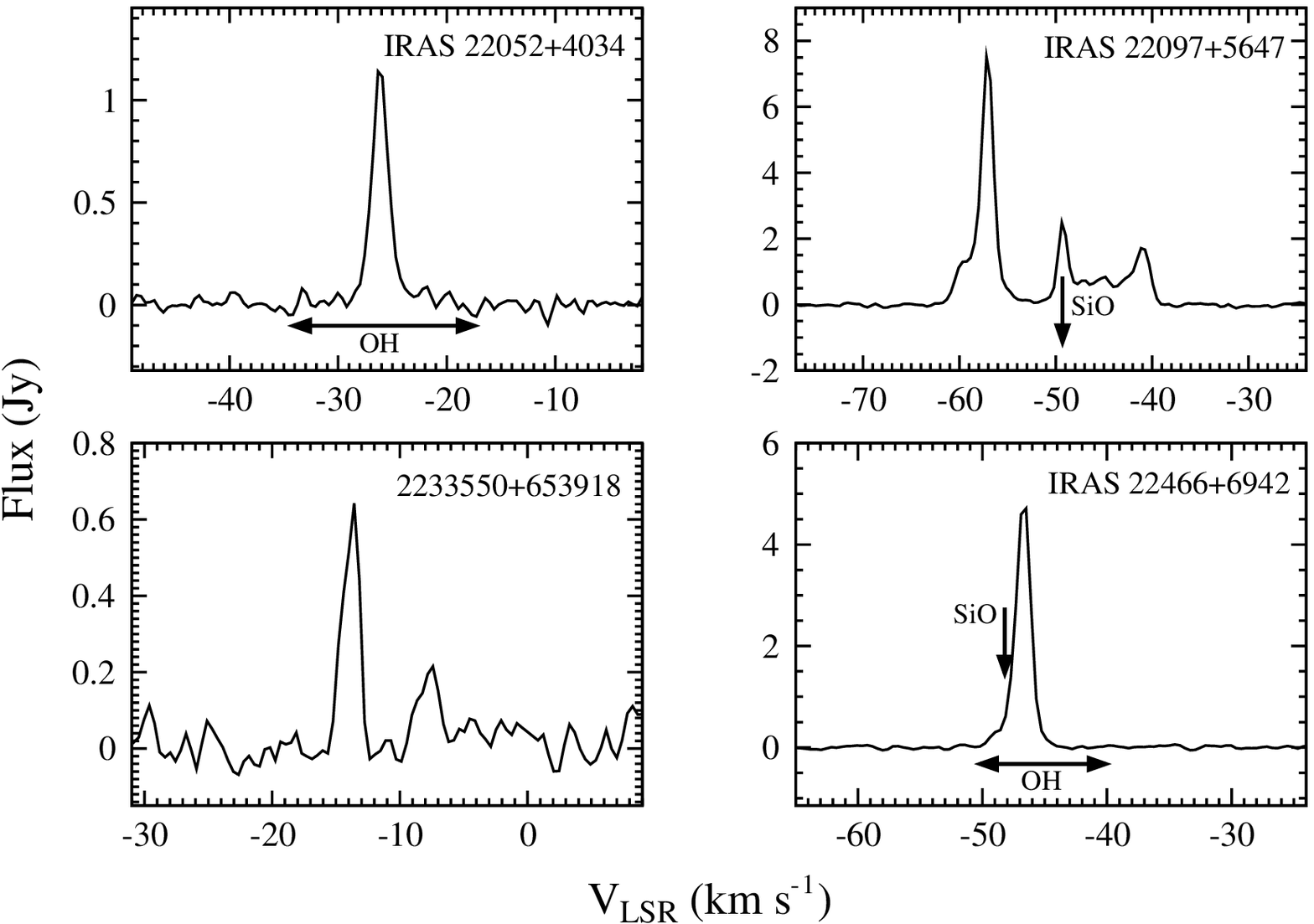}
   \caption{\it Continued}
\end{figure}
%%%%%%%%%%%%%%%%%%%% Figure 3 Ends %%%%%%%%%%%%%%%%%%%%%%%%%

%%%%%%%%%%%%%%%%%%%%%%% Figure 4 %%%%%%%%%%%%%%%%%%%%%%%%%%%
\begin{figure}[ht]
   \centering
   \includegraphics[scale=0.9]{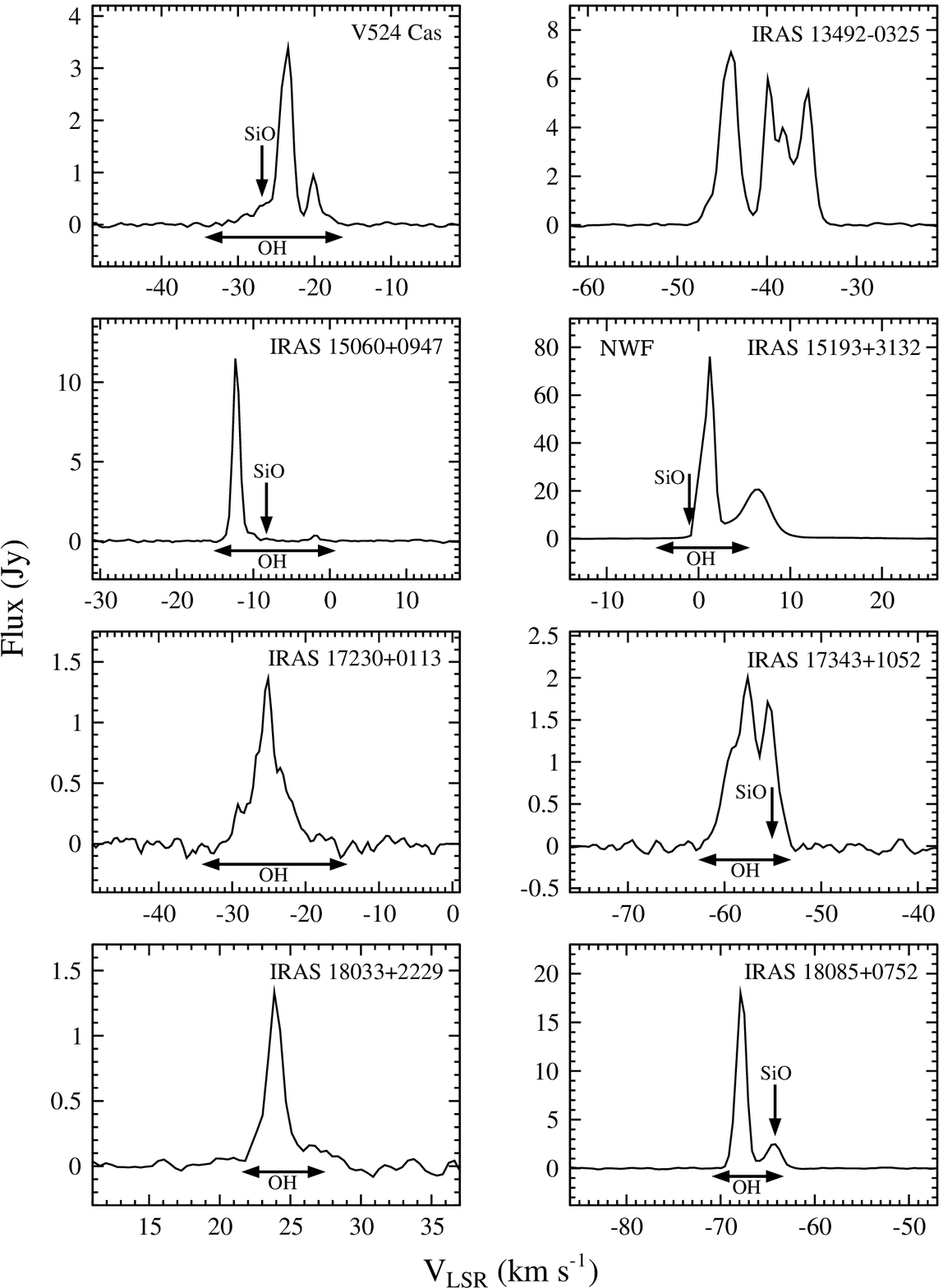}
   \caption{Spectra of revisited known H$_{2}$O maser detections.
            Some of them with new components detected, see
            Section~\ref{ssec:vel} and Appendix~\ref{app:oth}. 
            The new ``low-velocity'' water fountain 
            candidate (IRAS~15193$+$3132) is labeled as ``NWF''.}
   \label{fig:known_sp}
\end{figure}

\begin{figure}[ht]
   \figurenum{4}
   \centering
   \includegraphics[scale=0.9]{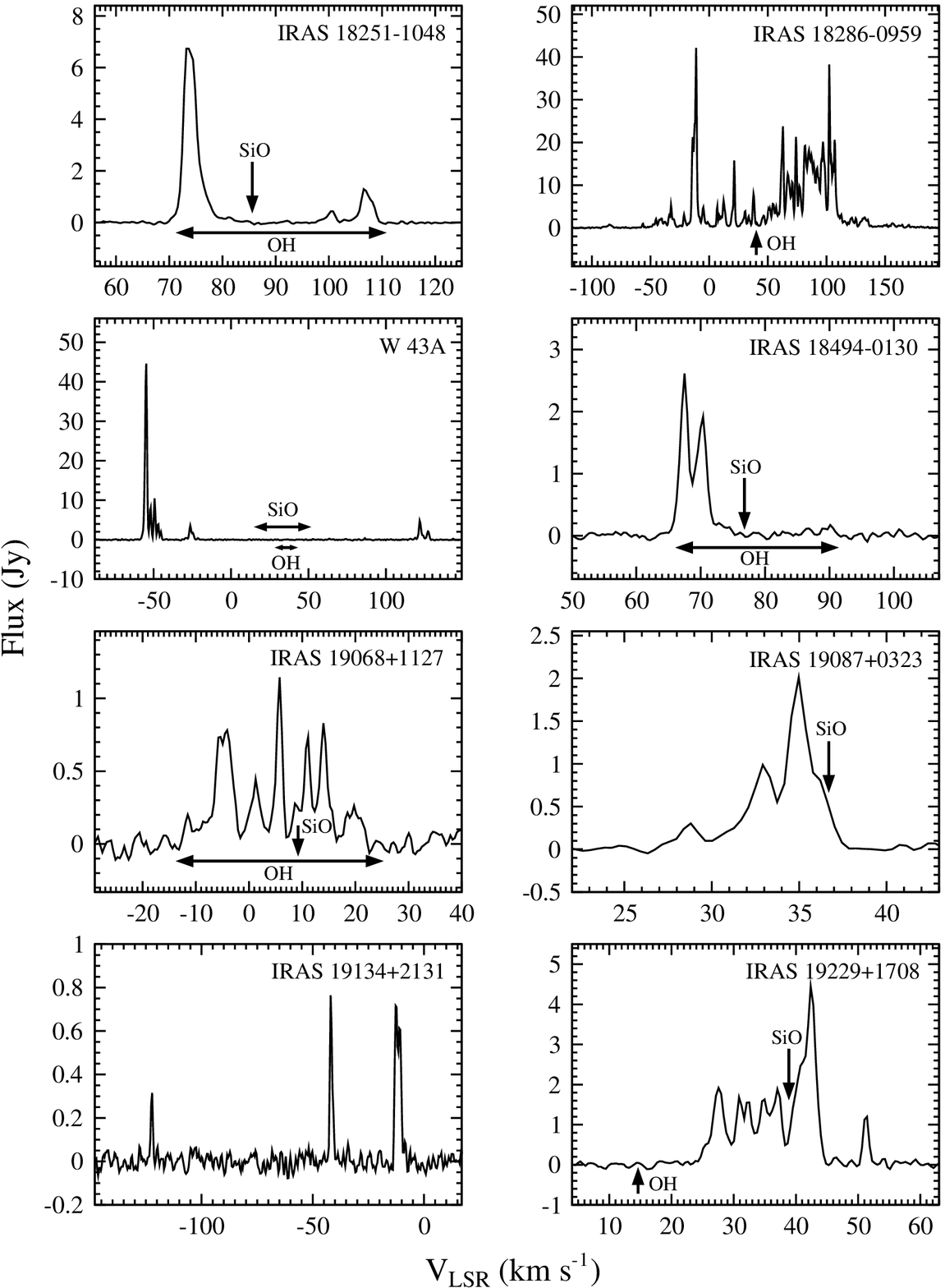}
   \caption{\it Continued}
\end{figure}

\begin{figure}[ht]
   \figurenum{4}
   \centering
   \includegraphics[scale=0.9]{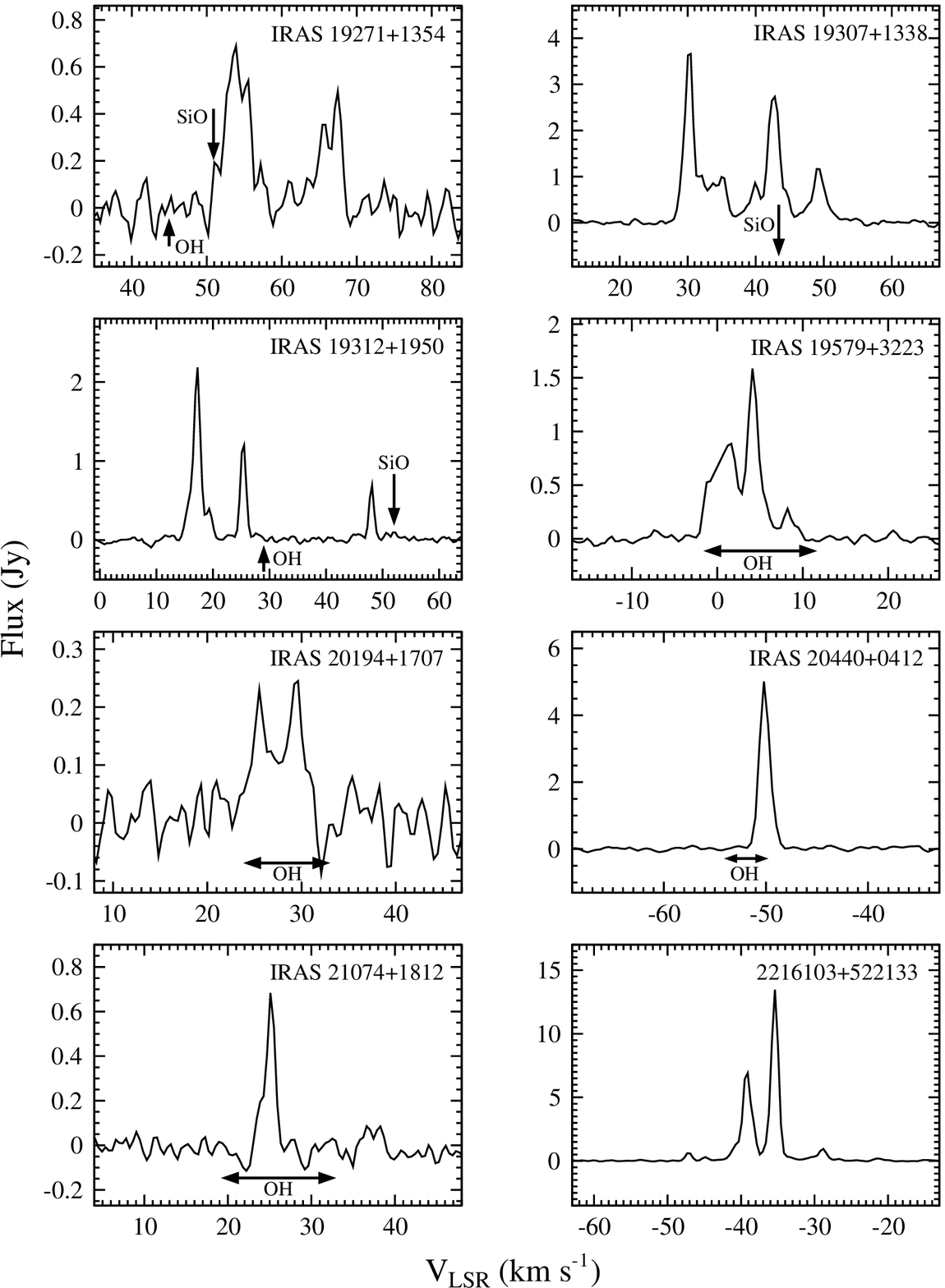}
   \caption{\it Continued}
\end{figure}

\begin{figure}[ht]
   \figurenum{4}
   \centering
   \includegraphics[scale=0.9]{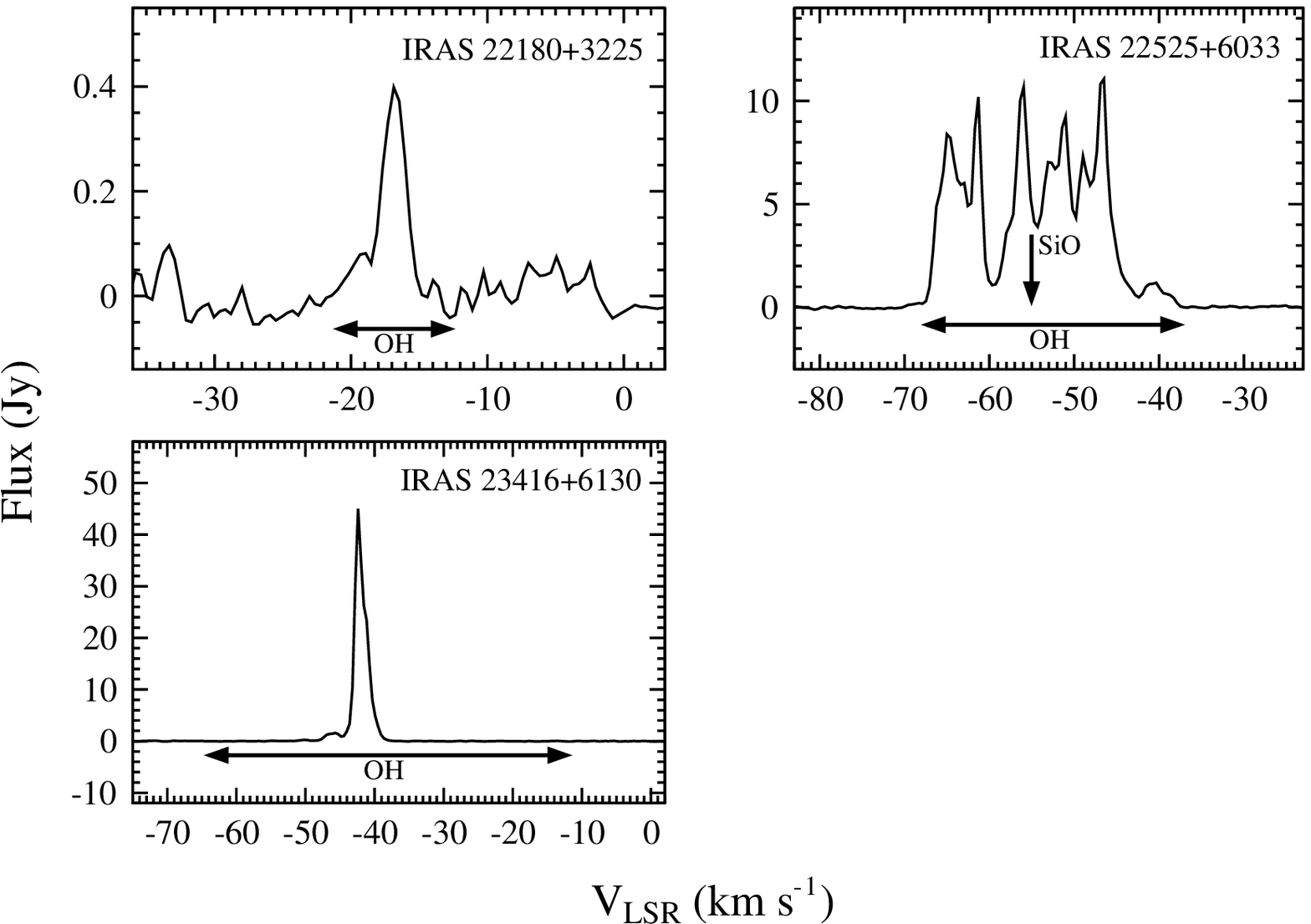}
   \caption{\it Continued}
\end{figure}

%%%%%%%%%%%%%%%%%%%% Figure 4 Ends %%%%%%%%%%%%%%%%%%%%%%%%%

%%%%%%%%%%%%%%%%%%%%%% Figure 5 %%%%%%%%%%%%%%%%%%%%%%%%%%%%

%%%%%%%%%%%%%%%%%%%% Figure 5 Ends %%%%%%%%%%%%%%%%%%%%%%%%%

%%%%%%% My figures End %%%%%%%%%%%%%

%% This figure uses \includegraphics to scale and rotate the still frame

%% If you are not including electonic art with your submission, you may
%% mark up your captions using the \figcaption command. See the
%% No more than seven \figcaption commands are allowed per page,
%% so if you have more than seven captions, insert a \clearpage

%% Tables should be submitted one per page, so put a \clearpage before

%% Two options are available to the author for producing tables:  the
%% deluxetable environment provided by the AASTeX package or the LaTeX
%% table environment.  Use of deluxetable is preferred.
%%

%% Three table samples follow, two marked up in the deluxetable environment,
%% one marked up as a LaTeX table.

\clearpage

%%%%% Table 1 %%%%%
\begin{deluxetable}{lrrrrrrrrcrrcl}
\rotate
\tablewidth{0pt}
\tabletypesize{\scriptsize}
\tablecaption
{Parameters of the observing targets. Non-detections are denoted by `N', 
and a blank entry means there were no corresponding observations carried out 
before.\label{tab:objs}}
\tablehead{

%obj_name|RA(observation)|Dec(observation)|IRAS_C12|IRAS_C23|AKARI_C12|AKARI_C23|OH(b_vel)|OH(r_vel)|OH_ref|SiO(v=1_vlsr)|SiO(v=2_vlsr)|SiO_ref|selection_criteria|

   \colhead{Object} &
   \colhead{R.A.\tablenotemark{a}} &
   \colhead{Decl.\tablenotemark{a}} &
   \colhead{IC$_{12}$\tablenotemark{b}} &
   \colhead{IC$_{23}$\tablenotemark{b}} &
   \colhead{AC$_{12}$\tablenotemark{c}} &
   \colhead{AC$_{23}$\tablenotemark{c}} &
   \colhead{OH $V_{\rm b,p}$\tablenotemark{d}} &
   \colhead{OH $V_{\rm r,p}$\tablenotemark{d}} &
   \colhead{Ref. 1\tablenotemark{e}} &
   \colhead{SiO \vlsr\ 1\tablenotemark{f}} &
   \colhead{SiO \vlsr\ 2\tablenotemark{f}} &
   \colhead{Ref. 2\tablenotemark{g}} &
   \colhead{Cat.\tablenotemark{h}} \\

   &
   &
   &
   &
   &
   &
   &
   \colhead{(\kms)} &
   \colhead{(\kms)} &
   &
   \colhead{(\kms)} &
   \colhead{(\kms)} 
}
\startdata
IRAS~23575$+$2536 & 00 00 06.56 & $+$25 53 11.2 & $-0.94$ & $-2.08$ & $-0.83$ & $\cdots$ & N & N & $1$ & $-29.2$ & $-29.1$ & $2$ & e \\
IRAS~00170$+$6542 & 00 19 51.28 & $+$65 59 30.4 & $0.19$ & $-1.68$ & $0.26$ & $-1.84$ & $-65.0$ & $-37.6$ & $3$ & $-48.9$ & $-51.0$ & $4$ & d,e \\
V~524CAS        & 00 46 00.12 & $+$69 10 53.4 & $-0.36$ & $-1.78$ & $-0.09$ & $-2.57$ & $-34.8$ & $-16.5$ & $5$ & $-27.0$ & $\cdots$ & $6$ & c,e \\
IRAS~01572$+$5844 & 02 00 44.10 & $+$58 59 03.0 & $-0.50$ & $-1.86$ & $\cdots$ & $\cdots$ & $-19.8$ & $-2.9$ & $7$ & $\cdots$ & $\cdots$ & $\cdots$ & d \\
IRAS~02547$+$1106 & 02 57 27.48 & $+$11 18 05.7 & $-0.09$ & $-1.71$ & $-0.31$ & $-2.22$ & $6.4$ & $25.4$ & $1$ & $\cdots$ & $\cdots$ & $\cdots$ & c,d \\
IRAS~03022$+$5409 & 03 05 52.91 & $+$54 20 53.9 & $-0.43$ & $-2.18$ & $-0.30$ & $\cdots$ & $\cdots$ & $\cdots$ & $\cdots$ & N & N & $8$ & d,e \\
IRAS~03206$+$6521 & 03 25 08.80 & $+$65 32 07.0 & $0.36$ & $-1.38$ & $0.71$ & $-1.26$ & $-47.0$ & $-28.0$ & $3$ & $-40$ & $-37.8$ & $6,9$ & b,d,e \\
IRAS~03461$+$6727 & 03 50 57.00 & $+$67 36 50.0 & $-0.74$ & $-1.60$ & $-0.64$ & $\cdots$ & $\cdots$ & $\cdots$ & $\cdots$ & $\cdots$ & $\cdots$ & $\cdots$ & f \\
IRAS~04209$+$4800 & 04 24 40.40 & $+$48 07 24.2 & $-0.28$ & $-1.91$ & $-0.57$ & $-2.09$ & $-26.4$ & $\cdots$ & $7$ & $-16.8$ & $-16.7$ & $9$ & c,d,e \\
IRAS~05131$+$4530 & 05 16 47.10 & $+$45 34 04.0 & $0.61$ & $-1.29$ & $0.61$ & $-1.43$ & $-42.8$ & $-22.9$ & $3$ & N & $-30.5$ & $8$ & c,d,e \\
IRAS~05284$+$1945 & 05 31 24.70 & $+$19 47 19.0 & $0.83$ & $-1.07$ & $0.51$ & $\cdots$ & $2.6$ & $27.0$ & $10$ & $14.2$ & $14.6$ & $9$ & d,e \\
IRAS~05506$+$2414 & 05 53 43.59 & $+$24 14 44.4 & $\cdots$ & $\cdots$ & $1.27$ & $1.20$ & $0.5$ & $10.1$ & $1$ & $\cdots$ & $\cdots$ & $\cdots$ & b \\
IRAS~05552$+$1720 & 05 58 07.51 & $+$17 20 58.5 & $\cdots$ & $\cdots$ & $-0.07$ & $\cdots$ & $31.8$ & $57.7$ & $1$ & $42.2$ & $45.1$ & $9$ & b,e \\
IRAS~06121$+$1221 & 06 14 59.40 & $+$12 20 16.0 & $\cdots$ & $\cdots$ & $\cdots$ & $\cdots$ & $82.3$ & $\cdots$ & $1$ & $\cdots$ & $\cdots$ & $\cdots$ & h \\
IRAS~06238$+$0904 & 06 26 37.26 & $+$09 02 14.9 & $-0.72$ & $-1.48$ & $-0.89$ & $-1.98$ & $26.0$ & $\cdots$ & $1$ & $\cdots$ & $\cdots$ & $\cdots$ & c,d \\
IRAS~06319$+$0415 & 06 34 37.63 & $+$04 12 42.8 & $1.70$ & $1.02$ & $1.68$ & $\cdots$ & $12.7$ & $\cdots$ & $1$ & N & N & $11$ & d \\
0759401$+$152312 & 07 59 40.13 & $+$15 23 12.4 & $\cdots$ & $1.61$ & $1.30$ & $1.68$ & $\cdots$ & $\cdots$ & $\cdots$ & $\cdots$ & $\cdots$ & $\cdots$ & b \\
IRAS~13492$-$0325 & 13 51 51.66 & $-$03 40 34.0 & $-0.82$ & $-1.89$ & $-0.74$ & $\cdots$ & N & N & $12$ & N & $\cdots$ & $13$ & d \\
IRAS~15060$+$0947 & 15 08 25.70 & $+$09 36 18.0 & $-0.29$ & $-2.06$ & $-0.17$ & $-2.36$ & $-16.3$ & $1.8$ & $1$ & $-8.2$ & $-8.3$ & $14$ & a,b,d,e \\
IRAS~15193$+$3132 & 15 21 23.30 & $+$31 22 02.0 & $-0.51$ & $-2.04$ & $-0.57$ & $-2.28$ & $-5.0$ & $5.5$ & $\cdots$\tablenotemark{i} & $-1.1$ & $-1.2$ & $15$ & a,d,e \\
IRAS~16030$-$0634 & 16 05 46.33 & $-$06 42 27.9 & $-0.64$ & $-1.73$ & $-0.46$ & $\cdots$ & $\cdots$ & $\cdots$ & $\cdots$ & $-6.0$ & $-6.4$ & $16$ & e \\
1611445$+$120416 & 16 11 44.55 & $+$12 04 16.6 & $\cdots$ & $\cdots$ & $3.56$ & $0.36$ & $\cdots$ & $\cdots$ & $\cdots$ & $\cdots$ & $\cdots$ & $\cdots$ & b \\
IRAS~16131$-$0216 & 16 15 47.66 & $-$02 23 31.9 & $-0.72$ & $-1.90$ & $-0.64$ & $\cdots$ & $\cdots$ & $\cdots$ & $\cdots$ & $60.3$ & $60.2$ & $14$ & d,e \\
1644295$+$234759 & 16 44 29.51 & $+$23 47 59.8 & $2.75$ & $0.32$ & $2.42$ & $0.83$ & N & N & $17$ & $\cdots$ & $\cdots$ & $\cdots$ & b \\
IRAS~17055$-$0216 & 17 08 10.20 & $-$02 20 21.0 & $-0.19$ & $-1.89$ & $-0.22$ & $\cdots$ & $\cdots$ & $\cdots$ & $\cdots$ & $-40.6$ & $-40.1$ & $16$ & d,e \\
IRAS~17132$-$0744 & 17 15 56.40 & $-$07 47 33.0 & $-0.50$ & $-2.03$ & $\cdots$ & $\cdots$ & $\cdots$ & $\cdots$ & $\cdots$ & $2.1$ & $3.7$ & $16$ & d,e \\
IRAS~17171$-$0843 & 17 19 53.45 & $-$08 46 59.7 & $-0.26$ & $-2.08$ & $-0.11$ & $\cdots$ & N & N & $18$ & $-15.5$ & $-15.6$ & $16$ & d,e \\
IRAS~17193$-$0601 & 17 22 02.30 & $-$06 04 13.0 & $-0.46$ & $-1.88$ & $-0.46$ & $\cdots$ & $\cdots$ & $\cdots$ & $\cdots$ & $-36.5$ & $-35.8$ & $16$ & d,e \\
IRAS~17230$+$0113 & 17 25 36.51 & $+$01 11 06.0 & $-0.17$ & $-1.84$ & $-0.13$ & $\cdots$ & $-34.6$ & $-14.4$ & $1$ & $\cdots$ & $\cdots$ & $\cdots$ & d \\
IRAS~17308$+$0822 & 17 33 13.90 & $+$08 20 41.0 & $0.05$ & $-1.55$ & $0.52$ & $\cdots$ & $1.9$ & $14.7$ & $1$ & $8.1$ & $8.0$ & $19$ & d,e \\
IRAS~17343$+$1052 & 17 36 44.50 & $+$10 51 05.0 & $-0.90$ & $-1.88$ & $-0.72$ & $\cdots$ & $-62.6$ & $-53.1$ & $1$ & $-55.3$ & $-55.4$ & $19$ & a,e \\
1744554$+$500239 & 17 44 55.44 & $+$50 02 39.5 & $3.70$ & $-0.21$ & $4.36$ & $0.58$ & N & N & $20$ & $\cdots$ & $\cdots$ & $\cdots$ & b \\
1749069$+$080610 & 17 49 06.91 & $+$08 06 10.2 & $1.11$ & $1.07$ & $1.18$ & $1.63$ & $\cdots$ & $\cdots$ & $\cdots$ & $\cdots$ & $\cdots$ & $\cdots$ & b \\
IRAS~17484$-$1511 & 17 51 20.38 & $-$15 12 26.6 & $0.74$ & $-0.95$ & $1.19$ & $-1.13$ & $77.0$ & $110.1$ & $10$ & N & N & $21$ & c,d \\
IRAS~17531$-$0940 & 17 55 53.10 & $-$09 41 24.0 & $0.73$ & $-1.14$ & $0.74$ & $\cdots$ & $18.6$ & $43.6$ & $3$ & $\cdots$ & $\cdots$ & $\cdots$ & d \\
1758333$+$663759 & 17 58 33.39 & $+$66 37 59.9 & $\cdots$ & $\cdots$ & $2.54$ & $0.78$ & N & N & $22$ & $\cdots$ & $\cdots$ & $\cdots$ & b \\
1800071$+$663654 & 18 00 07.14 & $+$66 36 54.3 & $\cdots$ & $\cdots$ & $1.93$ & $1.38$ & $\cdots$ & $\cdots$ & $\cdots$ & $\cdots$ & $\cdots$ & $\cdots$ & b \\
IRAS~18033$+$2229 & 18 05 26.60 & $+$22 30 04.0 & $\cdots$ & $\cdots$ & $-0.63$ & $\cdots$ & $21.9$ & $27.3$ & $1$ & $\cdots$ & $\cdots$ & $\cdots$ & a \\
IRAS~18050$-$0518 & 18 07 41.03 & $-$05 18 19.6 & $-0.16$ & $-2.06$ & $0.21$ & $-2.30$ & $-34.2$ & $-20.3$ & $18$ & $\cdots$ & $\cdots$ & $\cdots$ & c,d \\
IRAS~18056$-$1514 & 18 08 28.40 & $-$15 13 30.0 & $0.12$ & $-1.02$ & $0.43$ & $-0.96$ & $46.7$ & $72.1$ & $18$ & N & $59.7$ & $23$ & c,e \\
IRAS~18085$+$0752 & 18 10 58.50 & $+$07 53 09.0 & $-0.55$ & $-2.07$ & $-0.43$ & $\cdots$ & $-71.2$ & $-63.5$ & $1$ & N & $-64.1$ & $19$ & a,d,e \\
IRAS~18095$+$2704 & 18 11 30.67 & $+$27 05 15.5 & $1.12$ & $-1.64$ & $1.52$ & $-1.78$ & $-12.7$ & $0.9$ & $1$ & N & N & $21$ & c,d \\
1812063$+$065113 & 18 12 06.34 & $+$06 51 13.1 & $2.11$ & $-0.62$ & $2.47$ & $-0.19$ & N & N & $22$ & $\cdots$ & $\cdots$ & $\cdots$ & b \\
IRAS~18099$-$1449 & 18 12 47.37 & $-$14 48 50.0 & $-0.07$ & $\cdots$ & $-0.68$ & $\cdots$ & $\cdots$ & $\cdots$ & $\cdots$ & $1.7$ & $-0.5$ & $14$ & e \\
IRAS~18100$-$1250 & 18 12 50.49 & $-$12 49 44.8 & $-0.19$ & $\cdots$ & $-0.39$ & $\cdots$ & $\cdots$ & $\cdots$ & $\cdots$ & $80.2$ & $79.5$ & $14$ & e \\
IRAS~18117$-$1625 & 18 14 38.70 & $-$16 24 39.0 & $0.23$ & $\cdots$ & $0.04$ & $\cdots$ & $\cdots$ & $\cdots$ & $\cdots$ & $85.1$ & $84.8$ & $23$ & e \\
IRAS~18118$-$1615 & 18 14 41.35 & $-$16 14 03.0 & $-0.10$ & $\cdots$ & $0.47$ & $\cdots$ & $\cdots$ & $\cdots$ & $\cdots$ & $185.6$ & $184.7$ & $14$ & e \\
IRAS~18123$+$0511 & 18 14 49.41 & $+$05 12 55.2 & $0.03$ & $-1.04$ & $0.15$ & $-1.23$ & $85.0$ & $\cdots$ & $1$ & $\cdots$ & $\cdots$ & $\cdots$ & c,d \\
IRAS~18127$-$1516 & 18 15 39.90 & $-$15 15 13.0 & $0.06$ & $\cdots$ & $\cdots$ & $\cdots$ & $\cdots$ & $\cdots$ & $\cdots$ & $4.2$ & $1.0$ & $24$ & e \\
OH~15.7$+$0.8     & 18 16 25.72 & $-$14 55 14.5 & $1.51$ & $0.26$ & $3.07$ & $0.74$ & $-15.5$ & $12.8$ & $25$ & $-4.6$ & $-5.2$ & $26$ & b,d,e \\
1817244$-$170623 & 18 17 24.44 & $-$17 06 23.1 & $3.61$ & $1.39$ & $3.47$ & $1.29$ & $\cdots$ & $\cdots$ & $\cdots$ & $\cdots$ & $\cdots$ & $\cdots$ & b \\
1817340$+$100903 & 18 17 34.09 & $+$10 09 03.7 & $3.12$ & $-0.37$ & $3.56$ & $0.15$ & $\cdots$ & $\cdots$ & $\cdots$ & $\cdots$ & $\cdots$ & $\cdots$ & b \\
IRAS~18156$+$0655 & 18 18 07.19 & $+$06 56 17.6 & $\cdots$ & $\cdots$ & $-0.73$ & $\cdots$ & N & N & $27$ & $37.4$ & $37.3$ & $28$ & e \\
OH~18.8$+$0.4     & 18 24 05.25 & $-$12 26 14.1 & $\cdots$ & $\cdots$ & $3.36$ & $1.47$ & $-2.5$ & $27.0$ & $25$ & $\cdots$ & $\cdots$ & $\cdots$ & b \\
IRAS~18237$+$2150 & 18 25 51.04 & $+$21 52 14.4 & $\cdots$ & $\cdots$ & $-0.05$ & $\cdots$ & $31.8$ & $36.3$ & $1$ & $\cdots$ & $\cdots$ & $\cdots$ & h \\
IRAS~18236$-$0447 & 18 26 19.80 & $-$04 45 46.7 & $0.99$ & $-0.02$ & $1.70$ & $-0.05$ & $-36.9$ & $-14.8$ & $10$ & $\cdots$ & $\cdots$ & $\cdots$ & b,d \\
1827092$+$011427 & 18 27 09.27 & $+$01 14 28.0 & $2.35$ & $1.21$ & $1.75$ & $1.24$ & $\cdots$ & $\cdots$ & $\cdots$ & $\cdots$ & $\cdots$ & $\cdots$ & b \\
IRAS~18245$-$0552 & 18 27 12.00 & $-$05 51 01.1 & $-0.08$ & $-1.50$ & $0.11$ & $-1.92$ & $-15.3$ & $16.1$ & $3$ & $-2.5$ & $1.1$ & $14$ & c,d \\
IRAS~18251$-$1048 & 18 27 56.30 & $-$10 46 58.0 & $0.75$ & $\cdots$ & $0.80$ & $\cdots$ & $71.3$ & $110.6$ & $25$ & $85.4$ & $86.8$ & $26$ & e \\
OH~20.1$-$0.1     & 18 28 11.00 & $-$11 28 30.0 & $\cdots$ & $\cdots$ & $\cdots$ & $\cdots$ & $\cdots$ & $\cdots$ & $\cdots$ & $\cdots$ & $\cdots$ & $\cdots$ & g \\
1829161$+$001822 & 18 29 16.18 & $+$00 18 22.3 & $\cdots$ & $1.11$ & $0.62$ & $1.04$ & $\cdots$ & $\cdots$ & $\cdots$ & $\cdots$ & $\cdots$ & $\cdots$ & b \\
1829553$+$004939 & 18 29 55.33 & $+$00 49 39.5 & $0.86$ & $1.06$ & $0.60$ & $1.58$ & $\cdots$ & $\cdots$ & $\cdots$ & $\cdots$ & $\cdots$ & $\cdots$ & b,d \\
1830061$+$004233 & 18 30 06.17 & $+$00 42 33.6 & $\cdots$ & $\cdots$ & $0.58$ & $0.70$ & $\cdots$ & $\cdots$ & $\cdots$ & $\cdots$ & $\cdots$ & $\cdots$ & b \\
IRAS~18273$-$0738 & 18 30 06.99 & $-$07 36 50.9 & $0.11$ & $-1.54$ & $-0.29$ & $-0.65$ & $-14.0$ & $14.5$ & $3$ & $-1.7$ & $0.1$ & $24$ & c,d \\
IRAS~18286$-$0959 & 18 31 22.93 & $-$09 57 21.7 & $\cdots$ & $\cdots$ & $0.80$ & $0.13$ & $39.5$ & $\cdots$ & $29$ & N & N & $30$ & a,b \\
OH~16.3$-$3.0     & 18 31 31.51 & $-$16 08 46.5 & $\cdots$ & $\cdots$ & $0.83$ & $-0.62$ & $20.4$ & $36.3$ & $31$ & $\cdots$ & $\cdots$ & $\cdots$ & c \\
1833016$-$105011 & 18 33 01.67 & $-$10 50 11.0 & $0.74$ & $0.69$ & $0.64$ & $0.98$ & $\cdots$ & $\cdots$ & $\cdots$ & $\cdots$ & $\cdots$ & $\cdots$ & b,d \\
1834515$-$081820 & 18 34 51.60 & $-$08 18 21.0 & $1.61$ & $\cdots$ & $1.56$ & $1.83$ & $\cdots$ & $\cdots$ & $\cdots$ & $\cdots$ & $\cdots$ & $\cdots$ & b \\
1839230$-$055323 & 18 39 23.03 & $-$05 53 23.2 & $2.89$ & $\cdots$ & $3.38$ & $1.33$ & $\cdots$ & $\cdots$ & $\cdots$ & $\cdots$ & $\cdots$ & $\cdots$ & b \\
W~43A           & 18 47 41.16 & $-$01 45 11.7 & $1.61$ & $\cdots$ & $\cdots$ & $\cdots$ & $27.0$ & $41.0$ & $32$ & $20.1$\tablenotemark{j} & $20.0$ & $26$ & a,e \\
IRAS~18455$+$0448 & 18 48 02.30 & $+$04 51 30.5 & $0.32$ & $-0.91$ & $1.03$ & $-0.88$ & $27.0$ & $41.0$ & $33$ & $\cdots$ & $\cdots$ & $\cdots$ & c \\
IRAS~18476$+$0555 & 18 50 04.80 & $+$05 59 32.0 & $-0.16$ & $-1.79$ & $-0.18$ & $-1.82$ & $36.3$ & $64.5$ & $1$ & $\cdots$ & $\cdots$ & $\cdots$ & c \\
IRAS~18494$-$0130 & 18 52 01.45 & $-$01 26 46.4 & $0.17$ & $\cdots$ & $0.21$ & $\cdots$ & $66.0$ & $91.6$ & $25$ & N & $76.7$ & $14$ & e \\
IRAS~18501$+$1019 & 18 52 32.76 & $+$10 23 30.8 & $0.14$ & $-1.44$ & $0.37$ & $-1.61$ & $0.8$ & $27.7$ & $1$ & N & N & $34$ & c,d \\
IRAS~18517$+$0037 & 18 54 20.89 & $+$00 41 04.5 & $0.94$ & $-0.44$ & $0.99$ & $-0.14$ & $10.5$ & $42.7$ & $1$ & $31.4$ & $27.7$ & $26$ & b,d,e \\
1854250$+$004958 & 18 54 25.10 & $+$00 49 58.2 & $\cdots$ & $\cdots$ & $0.69$ & $1.41$ & $\cdots$ & $\cdots$ & $\cdots$ & $\cdots$ & $\cdots$ & $\cdots$ & b \\
OH~35.6$+$0.3     & 18 57 27.00 & $+$02 11 48.0 & $\cdots$ & $\cdots$ & $\cdots$ & $\cdots$ & $\cdots$ & $\cdots$ & $\cdots$ & $\cdots$ & $\cdots$ & $\cdots$ & g \\
IRAS~18578$+$0831 & 19 00 17.50 & $+$08 35 29.0 & $0.51$ & $-1.16$ & $0.30$ & $-1.71$ & $34.1$ & $63.6$ & $1$ & $49.3$ & $49.7$ & $26$ & c,d \\
IRAS~18587$+$0521 & 19 01 10.70 & $+$05 25 46.0 & $1.59$ & $2.44$ & $\cdots$ & $\cdots$ & $\cdots$ & $\cdots$ & $\cdots$ & N & N & $26$ & d \\
IRAS~18596$+$0605 & 19 02 04.69 & $+$06 10 09.5 & $0.03$ & $\cdots$ & $0.04$ & $\cdots$ & $\cdots$ & $\cdots$ & $\cdots$ & $68.2$ & $64.0$ & $14$ & e \\
IRAS~19010$+$0526 & 19 03 33.48 & $+$05 31 30.4 & $0.37$ & $\cdots$ & $0.49$ & $\cdots$ & $54.5$ & $77.2$ & $1$ & $41.7$ & $41.3$ & $14$ & e \\
IRAS~19017$+$0608 & 19 04 09.71 & $+$06 13 16.0 & $0.68$ & $-0.90$ & $0.64$ & $\cdots$ & $134.4$ & $163.5$ & $1$ & $148.3$ & $149.0$ & $26$ & d,e \\
IRAS~19024$+$1923 & 19 04 36.34 & $+$19 28 29.3 & $-0.19$ & $-1.98$ & $-0.05$ & $\cdots$ & $9.1$ & $45.4$ & $1$ & $27.7$ & $27.0$ & $34$ & d,e \\
IRAS~19023$+$0745 & 19 04 43.50 & $+$07 50 19.0 & $0.39$ & $-1.44$ & $0.32$ & $\cdots$ & $19.1$ & $47.2$ & $1$ & $31.3$ & $34.8$ & $34$ & d,e \\
IRAS~19027$+$0517 & 19 05 14.28 & $+$05 21 52.2 & $-0.01$ & $\cdots$ & $0.35$ & $\cdots$ & $\cdots$ & $\cdots$ & $\cdots$ & $33.0$ & $32.7$ & $14$ & e \\
IRAS~19029$+$0933 & 19 05 22.10 & $+$09 38 23.0 & $-0.16$ & $\cdots$ & $0.10$ & $\cdots$ & $42.7$ & $79.0$ & $35$ & $58.6$ & $61.3$ & $34$ & e \\
IRAS~19032$+$1715 & 19 05 28.67 & $+$17 20 12.2 & $-0.36$ & $-1.91$ & $-0.52$ & $\cdots$ & N & N & $36$ & $-17.9$ & $-16.9$ & $34$ & d,e \\
IRAS~19041$+$0952 & 19 06 31.13 & $+$09 57 17.0 & $0.05$ & $\cdots$ & $-0.05$ & $\cdots$ & $45.0$ & $\cdots$ & $1$ & $62.4$ & $68.4$ & $34$ & e \\
IRAS~19044$+$0833 & 19 06 49.40 & $+$08 37 50.0 & $0.06$ & $\cdots$ & $-0.12$ & $\cdots$ & N & N & $36$ & $32.4$ & $32.4$ & $34$ & e \\
IRAS~19047$+$1539 & 19 06 58.70 & $+$15 43 58.0 & $-0.07$ & $-1.81$ & $0.21$ & $\cdots$ & N & N & $38$ & $96.5$ & $96.7$ & $34$ & d,e \\
IRAS~19055$+$0225 & 19 08 03.18 & $+$02 30 30.2 & $0.37$ & $-1.70$ & $0.53$ & $-1.37$ & $3.7$ & $34.8$ & $18$ & $19.4$ & $19.4$ & $14$ & c,d,e \\
IRAS~19065$+$0832 & 19 08 58.53 & $+$08 37 48.1 & $1.02$ & $-0.54$ & $1.17$ & $0.15$ & $35.4$ & $70.8$ & $10$ & $52.0$ & $52.3$ & $26$ & b,d,e \\
IRAS~19068$+$1127 & 19 09 11.70 & $+$11 32 43.0 & $\cdots$ & $\cdots$ & $0.03$ & $\cdots$ & $-14.0$ & $25.0$ & $35$ & $9.3$ & $9.1$ & $34$ & a,e \\
IRAS~19071$+$0625 & 19 09 38.40 & $+$06 30 05.0 & $-0.06$ & $\cdots$ & $0.08$ & $\cdots$ & N & N & $35$ & $91.4$ & $91.0$ & $34$ & e \\
1909599$+$043708 & 19 09 59.97 & $+$04 37 08.1 & $1.82$ & $0.13$ & $1.71$ & $0.59$ & N & N & $17$ & $\cdots$ & $\cdots$ & $\cdots$ & b,d \\
IRAS~19079$+$1143 & 19 10 19.50 & $+$11 49 04.0 & $-0.09$ & $\cdots$ & $0.07$ & $\cdots$ & $24.5$ & $55.4$ & $35$ & $39.6$ & $37.6$ & $34$ & e \\
1910544$+$012444 & 19 10 54.50 & $+$01 24 44.2 & $2.06$ & $-0.09$ & $2.68$ & $0.10$ & N & N & $37$ & $\cdots$ & $\cdots$ & $\cdots$ & b \\
IRAS~19085$+$1038 & 19 10 57.20 & $+$10 43 38.0 & $\cdots$ & $\cdots$ & $\cdots$ & $\cdots$ & $\cdots$ & $\cdots$ & $\cdots$ & $\cdots$ & $\cdots$ & $\cdots$ & g \\
IRAS~19087$+$1413 & 19 11 05.40 & $+$14 18 20.0 & $-0.11$ & $-1.84$ & $-0.11$ & $\cdots$ & N & N & $18$ & $36.6$ & $36.4$ & $34$ & d,e \\
IRAS~19087$+$0323 & 19 11 16.97 & $+$03 28 24.2 & $0.30$ & $-1.60$ & $0.21$ & $-2.17$ & $6.4$ & $38.2$ & $35$ & $\cdots$ & $\cdots$ & $\cdots$ & c,d \\
IRAS~19114$+$0002 & 19 13 58.65 & $+$00 07 30.4 & $3.29$ & $-0.25$ & $4.24$ & $0.56$ & $73.0$ & $127.0$ & $39$ & N & N & $21$ & b \\
IRAS~19117$+$1107 & 19 14 19.60 & $+$11 10 35.0 & $\cdots$ & $\cdots$ & $0.65$ & $\cdots$ & $18.3$ & $52.6$ & $25$ & $33.5$ & $\cdots$ & $13$ & e \\
OH~45.5$+$0.0     & 19 14 24.00 & $+$11 09 24.0 & $\cdots$ & $\cdots$ & $\cdots$ & $\cdots$ & $18.0$ & $52.6$ & $32$ & $\cdots$ & $\cdots$ & $\cdots$ & g \\
1914408$+$114449 & 19 14 40.83 & $+$11 44 49.4 & $\cdots$ & $\cdots$ & $1.21$ & $1.15$ & $\cdots$ & $\cdots$ & $\cdots$ & $\cdots$ & $\cdots$ & $\cdots$ & b \\
IRAS~19134$+$2131 & 19 15 35.19 & $+$21 36 33.6 & $1.22$ & $-0.65$ & $2.35$ & $-0.24$ & N & N & $10$ & N & N & $26$ & a,d \\
IRAS~19201$+$2101 & 19 22 17.21 & $+$21 07 24.8 & $\cdots$ & $\cdots$ & $0.74$ & $-1.13$ & $41.9$ & $68.7$ & $36$ & $\cdots$ & $\cdots$ & $\cdots$ & c \\
IRAS~19231$+$3555 & 19 24 59.07 & $+$36 01 42.4 & $-0.23$ & $-2.09$ & $-0.10$ & $-2.51$ & $-34.8$ & $-13.9$ & $40$ & $-23.5$ & $-23.3$ & $34$ & c,d,e \\
IRAS~19229$+$1708 & 19 25 12.50 & $+$17 14 50.0 & $-0.28$ & $-1.68$ & $0.03$ & $\cdots$ & $14.6$ & $\cdots$ & $41$ & $38.8$ & $45.1$ & $34$ & a,d,e \\
IRAS~19271$+$1354 & 19 29 30.20 & $+$14 00 49.0 & $-0.46$ & $-1.95$ & $-0.28$ & $\cdots$ & $45.0$ & $\cdots$ & $36$ & $51.0$ & $50.9$ & $34$ & a,d,e \\
IRAS~19283$+$1421 & 19 30 38.05 & $+$14 27 55.7 & $0.01$ & $-1.95$ & $-0.18$ & $-1.78$ & $-19.5$ & $8.2$ & $1$ & $-5.1$ & $-4.8$ & $34$ & c,d,e \\
OH~53.6$-$0.2     & 19 31 22.50 & $+$18 13 20.0 & $\cdots$ & $\cdots$ & $\cdots$ & $\cdots$ & $-2.4$ & $24.4$ & $25$ & $\cdots$ & $\cdots$ & $\cdots$ & h \\
IRAS~19295$+$2228 & 19 31 38.97 & $+$22 35 17.2 & $0.63$ & $\cdots$ & $0.67$ & $-1.63$ & $-86.0$ & $-58.0$ & $42$ & $-73.2$ & $-72.8$ & $26$ & c,e \\
1932551$+$141337 & 19 32 55.14 & $+$14 13 37.8 & $3.04$ & $-0.67$ & $3.37$ & $-0.13$ & N & N & $43$ & $\cdots$ & $\cdots$ & $\cdots$ & b \\
IRAS~19307$+$1338 & 19 33 01.74 & $+$13 44 42.0 & $-0.26$ & $-2.17$ & $0.02$ & $-2.43$ & N & N & $41$ & $43.5$ & $43.5$ & $34$ & c,d,e \\
IRAS~19309$+$2022 & 19 33 07.20 & $+$20 28 59.0 & $-0.45$ & $\cdots$ & $0.18$ & $\cdots$ & $98.9$ & $108.1$ & $18$ & $45.6$ & $45.3$ & $34$ & e \\
IRAS~19312$+$1950 & 19 33 24.30 & $+$19 56 55.0 & $1.24$ & $1.79$ & $1.08$ & $2.32$ & $29.0$ & $\cdots$ & $44$ & $51.9$ & $52.1$ & $45$ & e \\
IRAS~19315$+$1807 & 19 33 46.02 & $+$18 13 56.6 & $-0.21$ & $\cdots$ & $0.05$ & $\cdots$ & N & N & $36$ & N & N & $34$ & h \\
IRAS~19323$+$2103 & 19 34 28.70 & $+$21 10 29.0 & $-0.23$ & $\cdots$ & $-0.10$ & $\cdots$ & N & N & $18$ & $14.0$ & $11.8$ & $34$ & e \\
IRAS~19349$+$1657 & 19 37 13.60 & $+$17 03 49.0 & $-0.14$ & $-2.25$ & $\cdots$ & $\cdots$ & N & N & $38$ & $12.5$ & $12.5$ & $34$ & d,e \\
IRAS~19374$+$1626 & 19 39 39.17 & $+$16 33 41.1 & $\cdots$ & $\cdots$ & $0.88$ & $-0.71$ & $-43.0$ & $-16.3$ & $10$ & N & N & $26$ & b \\
IRAS~19374$+$0550 & 19 39 53.03 & $+$05 57 53.2 & $\cdots$ & $\cdots$ & $-0.54$ & $-2.39$ & $-32.3$ & $-0.9$ & $40$ & $-17.0$ & $-17.5$ & $34$ & c,e \\
IRAS~19393$+$2447 & 19 41 27.00 & $+$24 54 56.0 & $-0.23$ & $-1.65$ & $-0.03$ & $\cdots$ & N & N & $36$ & $18.8$ & $18.0$ & $34$ & d,e \\
IRAS~19395$+$1949 & 19 41 43.42 & $+$19 56 31.7 & $-0.18$ & $-1.67$ & $-0.28$ & $\cdots$ & $25.0$ & $\cdots$ & $41$ & $43.2$ & $42.8$ & $34$ & d,e \\
IRAS~19414$+$2237 & 19 43 34.00 & $+$22 44 59.0 & $-0.44$ & $-2.22$ & $-0.65$ & $\cdots$ & $10.9$ & $47.2$ & $35$ & $51.8$ & $47.8$ & $34$ & d,e \\
IRAS~19440$+$2251 & 19 46 08.80 & $+$22 59 24.0 & $0.62$ & $-0.94$ & $0.90$ & $\cdots$ & $-25.0$ & $9.8$ & $10$ & $-9.6$ & $-9.1$ & $26$ & d,e \\
IRAS~19464$+$3514 & 19 48 15.96 & $+$35 22 06.1 & $0.20$ & $-1.47$ & $0.20$ & $-1.80$ & $11.4$ & $35.4$ & $18$ & $23.2$ & $22.4$ & $34$ & c,d,e \\
1949296$+$312716 & 19 49 29.62 & $+$31 27 16.1 & $4.62$ & $0.42$ & $4.45$ & $1.16$ & N & N & $46$ & $\cdots$ & $\cdots$ & $\cdots$ & b \\
1952516$+$394326 & 19 52 51.64 & $+$39 43 26.1 & $0.92$ & $-0.50$ & $0.91$ & $-0.50$ & N & N & $10$ & N & N & $21$ & b,d \\
IRAS~19579$+$3223 & 19 59 51.30 & $+$32 32 09.0 & $-0.60$ & $-1.76$ & $-0.19$ & $\cdots$ & $-1.8$ & $11.9$ & $36$ & $\cdots$ & $\cdots$ & $\cdots$ & a,d \\
IRAS~19583$+$1323 & 20 00 39.20 & $+$13 31 36.0 & $-0.48$ & $-2.18$ & $-0.26$ & $\cdots$ & N & N & $36$ & $-24.9$ & $-24.7$ & $34$ & d,e \\
OH~70.3$+$1.6     & 20 01 55.00 & $+$33 34 24.0 & $\cdots$ & $\cdots$ & $\cdots$ & $\cdots$ & $\cdots$ & $\cdots$ & $\cdots$ & $\cdots$ & $\cdots$ & $\cdots$ & g \\
2001595$+$324733 & 20 01 59.56 & $+$32 47 33.0 & $1.69$ & $-0.94$ & $1.75$ & $-0.35$ & N & N & $37$ & $\cdots$ & $\cdots$ & $\cdots$ & b,d \\
IRAS~20010$+$2508 & 20 03 08.30 & $+$25 17 27.0 & $-0.93$ & $-1.88$ & $-0.75$ & $\cdots$ & $\cdots$ & $\cdots$ & $\cdots$ & $24.9$ & $25.4$ & $8$ & e \\
IRAS~20023$+$2855 & 20 04 20.82 & $+$29 04 06.5 & $1.11$ & $-0.33$ & $1.02$ & $-0.55$ & $-76.9$ & $-52.6$ & $25$ & $-66.2$ & $-65.2$ & $26$ & c,d,e \\
IRAS~20021$+$2156 & 20 04 17.30 & $+$22 04 59.0 & $-0.80$ & $-2.06$ & $-0.27$ & $\cdots$ & $\cdots$ & $\cdots$ & $\cdots$ & $23.43$ & N & $8$ & d,e \\
IRAS~20020$+$1739 & 20 04 21.63 & $+$17 48 34.6 & $-0.54$ & $-2.03$ & $-0.52$ & $\cdots$ & N & N & $36$ & $30.88$ & $32.37$ & $8$ & d,e \\
IRAS~20029$+$2759 & 20 05 00.30 & $+$28 08 00.0 & $-0.57$ & $-1.40$ & $\cdots$ & $\cdots$ & N & N & $36$ & $22.47$ & N & $8$ & e \\
2005300$+$325138 & 20 05 30.02 & $+$32 51 38.3 & $1.33$ & $-0.20$ & $1.51$ & $-0.12$ & N & N & $37$ & $\cdots$ & $\cdots$ & $\cdots$ & b,d \\
IRAS~20043$+$2653 & 20 06 22.82 & $+$27 02 10.6 & $0.93$ & $-0.79$ & $1.05$ & $-0.77$ & $-17.8$ & $8.7$ & $10$ & $-3.9$ & $-5.5$ & $26$ & c,d,e \\
2010236$+$462739 & 20 10 23.69 & $+$46 27 39.7 & $2.59$ & $0.31$ & $2.33$ & $0.60$ & $\cdots$ & $\cdots$ & $\cdots$ & $\cdots$ & $\cdots$ & $\cdots$ & b \\
2012428$+$195922 & 20 12 42.81 & $+$19 59 22.4 & $2.35$ & $0.14$ & $2.32$ & $0.62$ & $\cdots$ & $\cdots$ & $\cdots$ & $\cdots$ & $\cdots$ & $\cdots$ & b \\
2013579$+$293354 & 20 13 57.95 & $+$29 33 54.0 & $2.30$ & $0.04$ & $2.32$ & $0.78$ & N & N & $17$ & $\cdots$ & $\cdots$ & $\cdots$ & b \\
2015573$+$470534 & 20 15 57.33 & $+$47 05 34.5 & $\cdots$ & $\cdots$ & $2.97$ & $0.60$ & $\cdots$ & $\cdots$ & $\cdots$ & $\cdots$ & $\cdots$ & $\cdots$ & b \\
IRAS~20156$+$2130 & 20 17 48.90 & $+$21 40 04.0 & $-0.32$ & $-1.68$ & $-0.35$ & $\cdots$ & $6.7$ & $\cdots$ & $1$ & $21.2$ & $21.5$ & $34$ & d,e \\
IRAS~20181$+$2234 & 20 20 21.92 & $+$22 43 48.5 & $0.27$ & $-1.72$ & $0.27$ & $-2.17$ & $29.4$ & $51.7$ & $3$ & $40.8$ & $40.4$ & $34$ & c,d,e \\
2021328$+$371218 & 20 21 32.84 & $+$37 12 18.4 & $2.08$ & $2.20$ & $2.47$ & $1.92$ & $\cdots$ & $\cdots$ & $\cdots$ & $\cdots$ & $\cdots$ & $\cdots$ & b \\
2021388$+$373111 & 20 21 38.81 & $+$37 31 12.0 & $\cdots$ & $1.96$ & $2.03$ & $1.41$ & $\cdots$ & $\cdots$ & $\cdots$ & $\cdots$ & $\cdots$ & $\cdots$ & b \\
IRAS~20194$+$1707 & 20 21 42.70 & $+$17 17 18.0 & $\cdots$ & $\cdots$ & $-0.66$ & $\cdots$ & $23.7$ & $33.2$ & $1$ & $\cdots$ & $\cdots$ & $\cdots$ & a \\
IRAS~20215$+$6243 & 20 22 20.05 & $+$62 53 02.2 & $-0.60$ & $-2.11$ & $-0.56$ & $\cdots$ & $\cdots$ & $\cdots$ & $\cdots$ & $17.2$ & $18.9$ & $28$ & d,e \\
IRAS~20266$+$3856 & 20 28 30.00 & $+$39 06 57.0 & $\cdots$ & $-0.07$ & $2.92$ & $0.80$ & $-48.9$ & $-27.1$ & $25$ & $\cdots$ & $\cdots$ & $\cdots$ & b \\
IRAS~20305$+$6246 & 20 31 26.54 & $+$62 56 49.8 & $-1.02$ & $-2.02$ & $-0.89$ & $\cdots$ & $\cdots$ & $\cdots$ & $\cdots$ & $-14.9$ & $-14.5$ & $28$ & d,e \\
2032541$+$375128 & 20 32 54.11 & $+$37 51 28.8 & $2.03$ & $1.21$ & $1.82$ & $1.89$ & N & N & $17$ & $\cdots$ & $\cdots$ & $\cdots$ & b \\
2045540$+$675738 & 20 45 54.02 & $+$67 57 38.5 & $1.02$ & $0.43$ & $0.85$ & $0.99$ & N & N & $46$ & $\cdots$ & $\cdots$ & $\cdots$ & b,d \\
IRAS~20440$+$0412 & 20 46 33.20 & $+$04 23 35.0 & $-0.60$ & $-1.90$ & $-0.73$ & $\cdots$ & $-54.0$ & $-49.9$ & $41$ & N & $\cdots$ & $47$ & a,d \\
IRAS~20444$+$0540 & 20 46 53.80 & $+$05 51 28.5 & $-0.20$ & $-1.62$ & $-0.50$ & $-2.14$ & $53.9$ & $71.1$ & $3$ & $\cdots$ & $\cdots$ & $\cdots$ & c,d \\
2048166$+$342724 & 20 48 16.64 & $+$34 27 24.4 & $\cdots$ & $-0.13$ & $4.38$ & $0.29$ & $\cdots$ & $\cdots$ & $\cdots$ & $\cdots$ & $\cdots$ & $\cdots$ & b \\
IRAS~20479$+$5336 & 20 49 20.70 & $+$53 48 02.0 & $-0.05$ & $-2.02$ & $0.05$ & $\cdots$ & N & N & $18$ & $-56.9$ & $-57.8$ & $9$ & d,e \\
IRAS~20523$+$5302 & 20 53 48.01 & $+$53 13 58.7 & $-0.23$ & $-1.74$ & $-0.07$ & $\cdots$ & N & N & $18$ & $-65.1$ & $-65.0$ & $9$ & d,e \\
OH~85.4$+$0.1     & 20 53 37.98 & $+$44 58 07.4 & $\cdots$ & $\cdots$ & $0.91$ & $-0.65$ & $-35.7$ & $-8.5$ & $25$ & $-23.8$ & $-23.9$ & $48$ & c,e \\
IRAS~21000$+$8251 & 20 56 10.05 & $+$83 03 25.3 & $-0.78$ & $-1.83$ & $-0.67$ & $\cdots$ & N & N & $27$ & $30.3$ & $30.2$ & $2$ & d,e \\
IRAS~20549$+$5245 & 20 56 24.26 & $+$52 57 01.0 & $-0.26$ & $-1.35$ & $-0.14$ & $-1.54$ & $14.5$ & $\cdots$ & $7$ & $-1.0$ & $\cdots$ & $6$ & c,d,e \\
2058537$+$441528 & 20 58 53.71 & $+$44 15 28.6 & $\cdots$ & $\cdots$ & $0.67$ & $1.47$ & $\cdots$ & $\cdots$ & $\cdots$ & $\cdots$ & $\cdots$ & $\cdots$ & b \\
2058555$+$493112 & 20 58 55.58 & $+$49 31 12.4 & $1.01$ & $-0.14$ & $1.12$ & $0.20$ & N & N & $10$ & $\cdots$ & $\cdots$ & $\cdots$ & b,d \\
2059141$+$782304 & 20 59 14.14 & $+$78 23 04.3 & $0.63$ & $1.29$ & $0.52$ & $1.64$ & $\cdots$ & $\cdots$ & $\cdots$ & $\cdots$ & $\cdots$ & $\cdots$ & b \\
IRAS~21074$+$1812 & 21 09 46.60 & $+$18 24 50.0 & $-0.51$ & $-2.36$ & $-0.57$ & $\cdots$ & $19.1$ & $33.2$ & $36$ & $\cdots$ & $\cdots$ & $\cdots$ & d \\
2119074$+$461846 & 21 19 07.47 & $+$46 18 46.7 & $1.94$ & $-0.09$ & $1.29$ & $0.39$ & $\cdots$ & $\cdots$ & $\cdots$ & $\cdots$ & $\cdots$ & $\cdots$ & b,d \\
IRAS~21216$+$5536 & 21 23 09.23 & $+$55 49 14.8 & $-0.48$ & $-1.73$ & $-0.40$ & $\cdots$ & N & N & $18$ & $-8.4$ & $-10.2$ & $9$ & d,e \\
IRAS~21341$+$5101 & 21 35 52.40 & $+$51 14 42.0 & $-0.68$ & $-1.91$ & $-0.61$ & $\cdots$ & N & N & $49$ & $1.4$ & $-6.6$ & $9$ & d,e \\
IRAS~21453$+$5959 & 21 46 52.64 & $+$60 13 48.5 & $-0.27$ & $-1.72$ & $-0.30$ & $-2.07$ & N & N & $49$ & $-45.3$ & $-44.8$ & $9$ & b,d,e \\
IRAS~21509$+$6234 & 21 52 19.37 & $+$62 48 39.5 & $-0.55$ & $-1.81$ & $-0.41$ & $\cdots$ & $\cdots$ & $\cdots$ & $\cdots$ & $-69.1$ & $-69.6$ & $9$ & d,e \\
IRAS~21522$+$6018 & 21 53 46.10 & $+$60 32 14.2 & $-0.71$ & $-1.96$ & $-0.55$ & $\cdots$ & $\cdots$ & $\cdots$ & $\cdots$ & $22.4$ & $20.1$ & $9$ & d,e \\
2154144$+$565726 & 21 54 14.49 & $+$56 57 26.4 & $1.07$ & $-0.28$ & $1.55$ & $-0.50$ & N & N & $10$ & $\cdots$ & $\cdots$ & $\cdots$ & b,d \\
IRAS~21554$+$6204 & 21 56 58.32 & $+$62 18 45.6 & $0.85$ & $-1.06$ & $0.76$ & $-1.15$ & $-32.5$ & $-8.7$ & $10$ & $-18.5$ & $-22.1$ & $4$ & c,d,e \\
IRAS~21563$+$5630 & 21 58 01.30 & $+$56 44 49.6 & $-0.53$ & $-1.65$ & $-0.58$ & $-1.90$ & $\cdots$ & $\cdots$ & $\cdots$ & $-11.3$ & $-5.5$ & $9$ & b,e \\
2204124$+$530401 & 22 04 12.45 & $+$53 04 02.0 & $3.46$ & $-0.58$ & $3.53$ & $-0.17$ & $\cdots$ & $\cdots$ & $\cdots$ & $\cdots$ & $\cdots$ & $\cdots$ & b \\
IRAS~22036$+$5306 & 22 05 30.50 & $+$53 21 33.0 & $1.85$ & $0.91$ & $1.21$ & $\cdots$ & $-65.6$ & $-16.6$ & $17$ & $\cdots$ & $\cdots$ & $\cdots$ & d \\
IRAS~22052$+$4034 & 22 07 20.10 & $+$40 48 42.0 & $-0.49$ & $-2.09$ & $\cdots$ & $\cdots$ & $-34.9$ & $-16.9$ & $7$ & $\cdots$ & $\cdots$ & $\cdots$ & d \\
IRAS~22097$+$5647 & 22 11 31.88 & $+$57 02 17.4 & $-0.43$ & $-2.28$ & $-0.12$ & $-2.72$ & $\cdots$ & $\cdots$ & $\cdots$ & $-49.2$ & $-48.5$ & $45$ & c,d,e \\
IRAS~22103$+$5120 & 22 12 15.40 & $+$51 35 03.0 & $-0.50$ & $-1.82$ & $-0.41$ & $\cdots$ & $-57.7$ & $\cdots$ & $7$ & $-50.4$ & $-50.3$ & $9$ & d,e \\
2216103$+$522133 & 22 16 10.39 & $+$52 21 33.2 & $1.62$ & $0.67$ & $1.85$ & $1.19$ & N & N & $17$ & $\cdots$ & $\cdots$ & $\cdots$ & b,d \\
2219055$+$613616 & 22 19 05.52 & $+$61 36 16.1 & $0.75$ & $0.82$ & $0.63$ & $1.34$ & $\cdots$ & $\cdots$ & $\cdots$ & $\cdots$ & $\cdots$ & $\cdots$ & b,d \\
2219520$+$633532 & 22 19 52.05 & $+$63 35 32.4 & $0.65$ & $\cdots$ & $0.68$ & $1.18$ & $\cdots$ & $\cdots$ & $\cdots$ & $\cdots$ & $\cdots$ & $\cdots$ & b \\
IRAS~22180$+$3225 & 22 20 20.12 & $+$32 40 27.3 & $\cdots$ & $\cdots$ & $-0.56$ & $\cdots$ & $-21.2$ & $-12.6$ & $36$ & $\cdots$ & $\cdots$ & $\cdots$ & h \\
2223557$+$505800 & 22 23 55.73 & $+$50 58 00.2 & $2.32$ & $0.37$ & $2.01$ & $0.81$ & N & N & $22$ & $\cdots$ & $\cdots$ & $\cdots$ & b \\
2224314$+$434310 & 22 24 31.44 & $+$43 43 10.9 & $\cdots$ & $\cdots$ & $2.98$ & $0.40$ & N & N & $43$ & $\cdots$ & $\cdots$ & $\cdots$ & b \\
2233550$+$653918 & 22 33 55.02 & $+$65 39 18.5 & $\cdots$ & $\cdots$ & $1.12$ & $1.73$ & $\cdots$ & $\cdots$ & $\cdots$ & $\cdots$ & $\cdots$ & $\cdots$ & b \\
2235235$+$751708 & 22 35 23.58 & $+$75 17 08.0 & $1.80$ & $1.01$ & $0.78$ & $0.94$ & N & N & $46$ & $\cdots$ & $\cdots$ & $\cdots$ & b \\
IRAS~22345$+$5809 & 22 36 27.70 & $+$58 25 31.0 & $-0.36$ & $-2.14$ & $-0.28$ & $-2.09$ & $-29.1$ & $-21.5$ & $40$ & $\cdots$ & $\cdots$ & $\cdots$ & c,d \\
IRAS~22394$+$6930 & 22 40 59.80 & $+$69 46 14.7 & $-0.65$ & $-1.78$ & $-0.63$ & $\cdots$ & $\cdots$ & $\cdots$ & $\cdots$ & $-42.5$ & $-39.8$ & $8$ & d,e \\
IRAS~22394$+$5623 & 22 41 27.10 & $+$56 39 08.0 & $-0.64$ & $-1.68$ & $-0.71$ & $\cdots$ & $\cdots$ & $\cdots$ & $\cdots$ & $-26.4$ & $-26.3$ & $9$ & d,e \\
IRAS~22466$+$6942 & 22 48 14.03 & $+$69 58 28.5 & $-0.47$ & $-1.98$ & $-0.47$ & $\cdots$ & $-50.8$ & $-39.6$ & $18$ & $-48.2$ & $-47.9$ & $9$ & d,e \\
2251389$+$515042 & 22 51 38.97 & $+$51 50 42.7 & $3.40$ & $-0.59$ & $3.60$ & $-0.22$ & $\cdots$ & $\cdots$ & $\cdots$ & $\cdots$ & $\cdots$ & $\cdots$ & b \\
IRAS~22517$+$2223 & 22 54 12.00 & $+$22 39 34.0 & $-0.50$ & $-2.15$ & $-0.25$ & $\cdots$ & $-44.8$ & $-43.0$ & $36$ & $\cdots$ & $\cdots$ & $\cdots$ & d \\
IRAS~22525$+$6033 & 22 54 31.90 & $+$60 49 38.0 & $-0.21$ & $-1.84$ & $-0.02$ & $-2.21$ & $-68.0$ & $-37.0$ & $\cdots$\tablenotemark{i} & $-55.5$ & $-55.0$ & $2$ & a,c,d,e \\
2259184$+$662547 & 22 59 18.41 & $+$66 25 47.8 & $1.29$ & $-0.39$ & $1.24$ & $-0.00$ & N & N & $37$ & N & N & $4$ & b,d,e \\
2310320$+$673939 & 23 10 32.00 & $+$67 39 40.0 & $1.55$ & $1.41$ & $1.47$ & $1.97$ & $\cdots$ & $\cdots$ & $\cdots$ & $\cdots$ & $\cdots$ & $\cdots$ & b \\
2312291$+$612534 & 23 12 29.16 & $+$61 25 34.1 & $1.27$ & $0.45$ & $1.69$ & $1.53$ & $\cdots$ & $\cdots$ & $\cdots$ & $\cdots$ & $\cdots$ & $\cdots$ & b,d \\
2332448$+$620348 & 23 32 44.84 & $+$62 03 48.9 & $1.79$ & $-0.87$ & $1.96$ & $-0.42$ & N & N & $37$ & $\cdots$ & $\cdots$ & $\cdots$ & b,d \\
IRAS~23352$+$5834 & 23 37 40.00 & $+$58 50 47.0 & $-0.53$ & $-1.67$ & $-1.01$ & $\cdots$ & $-94.1$ & $-86.0$ & $7$ & $\cdots$ & $\cdots$ & $\cdots$ & d \\
IRAS~23361$+$6437 & 23 38 27.10 & $+$64 54 37.0 & $0.09$ & $-1.35$ & $0.07$ & $\cdots$ & $54.4$ & $62.6$ & $18$ & N & N & $8$ & d \\
IRAS~23416$+$6130 & 23 44 03.27 & $+$61 47 22.0 & $0.07$ & $-1.54$ & $0.38$ & $-1.59$ & $-65.0$ & $-12.0$ & $3$ & N & N & $2$ & c,d \\
IRAS~23489$+$6235 & 23 51 27.28 & $+$62 51 47.1 & $-1.06$ & $\cdots$ & $-1.44$ & $\cdots$ & $\cdots$ & $\cdots$ & $\cdots$ & $-41.0$ & $\cdots$ & $6$ & e \\
IRAS~23554$+$5612 & 23 58 01.32 & $+$56 29 13.4 & $-0.63$ & $-1.29$ & $-0.57$ & $-2.19$ & $\cdots$ & $\cdots$ & $\cdots$ & $8.0$ & $\cdots$ & $6$ & c,e \\
IRAS~23561$+$6037 & 23 58 38.70 & $+$60 53 48.0 & $-0.81$ & $-1.22$ & $-0.78$ & $\cdots$ & $\cdots$ & $\cdots$ & $\cdots$ & $-54.9$ & N & $8$ & e \\
\enddata

\tablenotetext{a}{J2000.0.}
\tablenotetext{b}{IC$_{12}$ and IC$_{23}$ represent the IRAS [12]$-$[25] and
                  [25]$-$[60] colors, respectively.}
\tablenotetext{c}{AC$_{12}$ and AC$_{23}$ represent the AKARI [09]$-$[18] and
                  [18]$-$[65] colors, respectively.}
\tablenotetext{d}{$V_{\rm b,p}$ and $V_{\rm r,p}$ represent the \vlsr\ of the
                  blueshifted and redshifted peak of a double-peaked 1612~MHz
                  OH maser profile, respectively. For a single-peaked profile, 
                  the \vlsr\ is recorded as $V_{\rm b,p}$, no matter it is 
                  really ``blueshifted'' or not.}
\tablenotetext{e}{References for 1612~MHz OH maser velocities.}
\tablenotetext{f}{``SiO \vlsr\ 1'' and ``SiO \vlsr\ 2'' represent 
                  the \vlsr\ of the SiO maser peak in the $(v=1,J=1-0)$ and
                  $(v=2,J=1-0)$ transitions, respectively.}
\tablenotetext{g}{References for SiO maser velocities.}
\tablenotetext{h}{Category, from (a) to (g), of which the object belongs to.}
\tablenotetext{i}{From our unpublished data of an OH maser observation
                  conducted in year 2012, using the Effelsberg 100~m radio 
                  telescope.}
\tablenotetext{j}{\citet{imai05apj} found that W~43A actually has a biconical
                  flow traced by SiO emission. The spectral velocity range is 
                  from about 15--50\kms.}
{\bf References.} 
(1)~\citet{lewis94apjs}, (2)~\citet{cho96aas}, (3)~\citet{lesqueren92aas},
(4)~\citet{fujii01}, (5)~\citet{szymczak95mnras}, (6)~Pointing list of the
NRO~45~m telescope, (7)~\citet{sivagnanam90aa}, (8)~\citet{jiang99pasj},
(9)~\citet{jiang96apjs}, (10)~\citet{david93aa}, (11)~\citet{zapata09apj},
(12)~\citet{lewis95aas}, (13)~\citet{jewell91aa},
(14)~\citet{deguchi04pasj}, (15)~\citet{kim10apjs}, (16)~\citet{ita01aa},
(17)~\citet{hekkert91aa}, (18)~\citet{hekkert91aas}, (19)~\citet{deguchi10pasj},
(20)~\citet{szymczak04aa}, (21)~\citet{nyman98aas}, (22)~\citet{payne98apj},
(23)~\citet{deguchi00apjs}, (24)~\citet{izumiura99apjs}, 
(25)~\citet{engels07aa}, (26)~\citet{nakashima03pasj2}, 
(27)~\citet{sivagnanam88aas}, (28)~\citet{matsuura00pasj},
(29)~\citet{imai08evn}, (30)~\citet{deguchi07apj}, 
(31)~\citet{sevenster01aa}, (32)~\citet{hekkert89aas}, (33)~\citet{lewis01apj},
(34)~\citet{nakashima03pasj1}, (35)~\citet{eder88apjs}, 
(36)~\citet{chengalur93apjs}, (37)~\citet{hu94aas}, (38)~\citet{lewis87aj},
(39)~\citet{gledhill01mnras}, (40)~\citet{josselin98aas}, 
(41)~\citet{lewis90apj}, (42)~\citet{engels96aa}, (43)~\citet{likkel89apj},
(44)~\citet{nakashima11apj}, (45)~\citet{nakashima07apj}, 
(46)~\citet{hekkert96aas}, (47)~\citet{lepine78apj}, 
(48)~\citet{deguchi05pasj2}, (49)~\citet{galt89aj}.

\end{deluxetable}

%%%%% Table 1 ends %%%%%

%%%%% Table 2 %%%%%
\begin{deluxetable}{cccc}
\tablewidth{0pt}
%\tabletypesize{\scriptsize}
\tablecaption
{Number of H$_{2}$O maser detections in each category described in
Section~\ref{ssec:cate}.\label{tab:num}}
\tablehead{
   \colhead{Category} &
   \colhead{Objects Observed} &
   \colhead{New Masers} &
   \colhead{Known Masers} 
}
\startdata
a &  15 &  0 & 15 \\ 
b &  68 &  6 &  3 \\ 
c &  38 & 11 &  5 \\ 
d & 106 & 21 & 16 \\ 
e &  98 & 22 & 14 \\
f &   1 &  1 &  0 \\ 
g &   5 &  3 &  0 \\ 
h &   5 &  0 &  1 \\ 
\enddata
\end{deluxetable}
%%%%% Table 2 ends %%%%%

%%%%% Table 3 %%%%%
\begin{deluxetable}{lrrrrrrrrc}
\tablewidth{0pt}
\tabletypesize{\scriptsize}
\tablecaption
{Parameters of the H$_{2}$O maser detections.\label{tab:h2o}}
\tablehead{
   \colhead{Object} &
   \colhead{$V_{\rm b,p}$\tablenotemark{a}} &
   \colhead{$F_{\rm b,p}$\tablenotemark{a}} &
   \colhead{$V_{\rm r,p}$\tablenotemark{b}} &
   \colhead{$F_{\rm r,p}$\tablenotemark{b}} &
   \colhead{$V_{\rm b,e}$\tablenotemark{c}} &
   \colhead{$V_{\rm r,e}$\tablenotemark{c}} &
   \colhead{$I$\tablenotemark{d}} &
   \colhead{rms} &
   \colhead{Ref.\tablenotemark{e}} \\

   &
   \colhead{(\kms)} &
   \colhead{(Jy)} &
   \colhead{(\kms)} &
   \colhead{(Jy)} &
   \colhead{(\kms)} &
   \colhead{(\kms)} &
   \colhead{(Jy~\kms)} &
   \colhead{($10^{-2}$~Jy)}     
}
\startdata
IRAS~00170$+$6542 & $-$63.4 & 2.21 & $\cdots$ & $\cdots$ & $-$65 & $-$58 & 4.80 & 3.23 & new \\
V524~CAS        & $-$23.5 & 3.36 & $\cdots$ & $\cdots$ & $-$32 & $-$16 & 10.08 & 2.46 & 1 \\
IRAS~03461$+$6727 & $-$37.9 & 4.22 & $\cdots$ & $\cdots$ & $-$40 & $-$36 & 5.60 & 4.61 & new \\
IRAS~04209$+$4800 & $-$21.0 & 7.55 & $\cdots$ & $\cdots$ & $-$27 & $-$11 & 19.26 & 2.75 & new \\
IRAS~06319$+$0415 & 7.8 & 0.42 & $\cdots$ & $\cdots$ & 6 & 11 & 0.77 & 4.93 & new \\
IRAS~13492$-$0325 & $-$44.0 & 7.04 & $\cdots$ & $\cdots$ & $-$47 & $-$36 & 35.90 & 3.42 & 1 \\
IRAS~15060$+$0947 & $-$12.4 & 11.52 & $-$2.0 & 0.32 & $-$14 & $-$1 & 15.26 & 4.48 & 2 \\
IRAS~15193$+$3132 & 1.3 & 75.52 & 6.4 & 20.58 & 1 & 11 & 147.30 & 5.15 & 2 \\
IRAS~17132$-$0744 & 4.1 & 0.74 & 9.9 & 0.22 & 2 & 12 & 1.60 & 4.70 & new \\
IRAS~17171$-$0843 & $-$9.5 & 0.70 & $\cdots$ & $\cdots$ & $-$12 & $-$7 & 1.25 & 4.86 & new \\
IRAS~17230$+$0113 & $-$25.1 & 1.38 & $\cdots$ & $\cdots$ & $-$32 & $-$20 & 5.31 & 4.16 & 2 \\
IRAS~17343$+$1052 & $-$57.6 & 1.98 & $-$55.5 & 1.70 & $-$63 & $-$53 & 8.77 & 4.64 & 2 \\
IRAS~18033$+$2229 & 23.9 & 1.34 & $\cdots$ & $\cdots$ & 22 & 26 & 2.08 & 4.48 & 2 \\
IRAS~18050$-$0518 & $-$27.2 & 1.02 & $\cdots$ & $\cdots$ & $-$32 & $-$20 & 3.26 & 4.83 & new \\
IRAS~18056$-$1514 & 59.3 & 6.40 & $\cdots$ & $\cdots$ & 36 & 64 & 33.79 & 3.10 & new \\
IRAS~18085$+$0752 & $-$67.8 & 17.63 & $-$64.2 & 2.46 & $-$70 & $-$62 & 29.92 & 4.32 & 2 \\
1817244$-$170623 & 21.4 & 1.50 & $\cdots$ & $\cdots$ & 20 & 23 & 2.24 & 3.65 & new \\
IRAS~18251$-$1048 & 73.5 & 6.75 & 106.6 & 1.28 & 70 & 110 & 27.14 & 3.39 & 3 \\
OH~20.1$-$0.1     & 46.5 & 3.49 & $\cdots$ & $\cdots$ & 36 & 48 & 9.18 & 5.89 & new \\
IRAS~18286$-$0959 & $-$11.1 & 42.24 & $\cdots$ & $\cdots$ & $-$92 & 171 & 1020.93 & 5.31 & $4,5$ \\
OH~16.3$-$3.0     & 15.6 & 0.45 & 39.5 & 0.48 & 14 & 43 & 2.85 & 7.01 & new \\
1833016$-$105011 & 51.0 & 5.18 & $\cdots$ & $\cdots$ & 50 & 52 & 4.96 & 3.49 & new \\
W~43A           & $-$54.7 & 44.80 & 1.5 & 389.76 & $-$58 & 129 & 127.14 & 6.34 & 6 \\
IRAS~18455$+$0448 & 10.3 & 0.26 & 46.1 & 0.96 & 9 & 48 & 3.68 & 3.71 & new \\
IRAS~18476$+$0555 & 43.2 & 0.64 & $\cdots$ & $\cdots$ & 40 & 48 & 2.02 & 2.46 & new \\
IRAS~18494$-$0130 & 67.5 & 2.56 & 70.4 & 1.89 & 65 & 72 & 7.94 & 4.26 & 3 \\
IRAS~18578$+$0831 & 38.7 & 0.19 & $\cdots$ & $\cdots$ & 37 & 41 & 0.35 & 1.86 & new \\
IRAS~19029$+$0933 & 49.8 & 1.12 & 71.6 & 0.58 & 44 & 73 & 5.41 & 4.10 & new \\
IRAS~19032$+$1715 & $-$24.3 & 3.84 & $\cdots$ & $\cdots$ & $-$27 & $-$9 & 27.10 & 3.87 & new \\
IRAS~19065$+$0832 & 64.6 & 0.13 & $\cdots$ & $\cdots$ & 58 & 68 & 0.48 & 1.92 & new \\
IRAS~19068$+$1127 & 5.8 & 1.15 & $\cdots$ & $\cdots$ & $-$13 & 22 & 10.24 & 4.83 & 2 \\
IRAS~19087$+$1413 & 42.0 & 0.96 & $\cdots$ & $\cdots$ & 38 & 45 & 2.78 & 4.32 & new \\
IRAS~19087$+$0323 & 35.0 & 2.02 & $\cdots$ & $\cdots$ & 27 & 38 & 5.79 & 3.42 & 7 \\
OH~45.5$+$0.0     & 51.4 & 0.70 & 58.8 & 0.42 & 49 & 61 & 1.50 & 5.15 & new \\
IRAS~19134$+$2131 & $-$122.2 & 3.10 & $-$12.7 & 0.70 & $-$123 & $-$8 & 3.42 & 2.91 & 8 \\
IRAS~19229$+$1708 & 42.4 & 4.42 & $\cdots$ & $\cdots$ & 25 & 53 & 30.37 & 4.90 & 2 \\
IRAS~19271$+$1354 & 53.9 & 0.67 & 67.5 & 0.48 & 50 & 69 & 3.97 & 7.07 & 2 \\
IRAS~19295$+$2228 & $-$81.1 & 0.45 & $\cdots$ & $\cdots$ & $-$85 & $-$80 & 1.18 & 3.39 & new \\
IRAS~19307$+$1338 & 30.4 & 3.68 & $\cdots$ & $\cdots$ & 28 & 52 & 21.41 & 4.16 & 2 \\
IRAS~19312$+$1950 & 17.3 & 2.11 & 48.2 & 0.70 & 14 & 49 & 7.01 & 3.10 & $9,10$ \\
IRAS~19349$+$1657 & $-$1.2 & 0.45 & $\cdots$ & $\cdots$ & $-$3 & 3 & 1.18 & 4.29 & new \\
IRAS~19374$+$0550 & $-$23.0 & 2.91 & $-$6.6 & 0.48 & $-$29 & $-$3 & 11.94 & 4.90 & new \\
IRAS~19393$+$2447 & 4.9 & 0.90 & $\cdots$ & $\cdots$ & 3 & 17 & 4.51 & 2.59 & new \\
IRAS~19395$+$1949 & 34.2 & 0.80 & $\cdots$ & $\cdots$ & 28 & 56 & 4.35 & 4.48 & new \\
IRAS~19464$+$3514 & 13.0 & 0.06 & 33.2 & 0.83 & 12 & 35 & 1.38 & 1.44 & new \\
IRAS~19579$+$3223 & 4.1 & 1.57 & $\cdots$ & $\cdots$ & $-$2 & 10 & 6.24 & 2.98 & 2 \\
OH~70.3$+$1.6     & $-$43.2 & 0.19 & $-$19.8 & 1.57 & $-$45 & $-$18 & 2.24 & 5.63 & new \\
IRAS~20029$+$2759 & 17.7 & 0.38 & 24.3 & 3.39 & 15 & 28 & 8.38 & 4.77 & new \\
IRAS~20194$+$1707 & 25.5 & 0.22 & 29.6 & 0.26 & 23 & 32 & 1.06 & 4.13 & 2 \\
IRAS~20440$+$0412 & $-$50.2 & 4.83 & $\cdots$ & $\cdots$ & $-$52 & $-$48 & 6.50 & 4.06 & 2 \\
IRAS~21074$+$1812 & 25.1 & 0.77 & $\cdots$ & $\cdots$ & 23 & 27 & 0.96 & 3.71 & 7 \\
IRAS~21216$+$5536 & $-$9.1 & 2.30 & $\cdots$ & $\cdots$ & $-$20 & $-$4 & 11.74 & 4.03 & new \\
IRAS~21341$+$5101 & $-$6.2 & 2.98 & $\cdots$ & $\cdots$ & $-$8 & $-$6 & 2.78 & 4.16 & new \\
IRAS~21453$+$5959 & $-$37.0 & 1.12 & $\cdots$ & $\cdots$ & $-$60 & $-$27 & 6.14 & 4.00 & new \\
IRAS~21563$+$5630 & $-$6.6 & 0.61 & $\cdots$ & $\cdots$ & $-$7 & $-$5 & 0.77 & 4.26 & new \\
IRAS~22052$+$4034 & $-$26.3 & 1.15 & $\cdots$ & $\cdots$ & $-$29 & $-$23 & 2.34 & 3.30 & new \\
IRAS~22097$+$5647 & $-$57.2 & 7.52 & $\cdots$ & $\cdots$ & $-$62 & $-$39 & 26.98 & 3.78 & new \\
2216103$+$522133 & $-$35.4 & 13.47 & $\cdots$ & $\cdots$ & $-$48 & $-$28 & 32.64 & 3.42 & 2 \\
IRAS~22180$+$3225 & $-$16.8 & 0.38 & $\cdots$ & $\cdots$ & $-$20 & $-$15 & 0.90 & 3.33 & 7 \\
2233550$+$653918 & $-$13.6 & 0.64 & $-$7.4 & 0.22 & $-$16 & $-$6 & 1.38 & 3.90 & new \\
IRAS~22466$+$6942 & $-$46.5 & 4.70 & $\cdots$ & $\cdots$ & $-$50 & $-$44 & 7.94 & 3.30 & new \\
IRAS~22525$+$6033 & $-$46.5 & 10.94 & $\cdots$ & $\cdots$ & $-$68 & $-$38 & 137.15 & 4.61 & 2 \\
IRAS~23416$+$6130 & $-$45.7 & 1.60 & $-$42.4 & 44.80 & $-$48 & $-$39 & 89.50 & 3.26 & 11 \\
\enddata

\tablenotetext{a}{\vlsr\ and flux density of the blueshifted peak of a
                  double-peaked profile. For a single-peaked or irregular 
                  profile, the brightest peak is recorded in these two 
                  columns, no matter it is really ``blueshifted'' or not.}
\tablenotetext{b}{Same as above, but for the redshifted peak of a
                  double-peaked profile, if exist.}
\tablenotetext{c}{\vlsr\ of the two ends of the whole emission profile.
                  The cut-off is defined by the 3-$\sigma$ flux level.}
\tablenotetext{d}{Integrated flux of the whole emission profile.}
\tablenotetext{e}{References for known detections.}
{\bf References.}
(1)~\citet{comoretto90aas}, (2)~\citet{valdettaro01aa}, (3)~\citet{engels86aa}
(4)~\citet{deguchi07apj}, (5)~\citet{yung11apj}, (6)~\citet{imai02nature},
(7)~\citet{engels96aas}, (8)~\citet{imai07apj},
(9)~\citet{nakashima00pasj}, (10)~\citet{nakashima11apj},
(11)~\citet{shintani08pasj}.
\end{deluxetable}
%%%%% Table 3 ends %%%%%

%%%%% Table 4 %%%%%
\begin{deluxetable}{lr}
\tablewidth{0pt}
\tabletypesize{\scriptsize}
\tablecaption
{Parameters of the non-detections.\label{tab:non}}
\tablehead{

   \colhead{Object} &
   \colhead{rms} \\

   & 
   \colhead{($10^{-2}$~Jy)}
}
\startdata
IRAS~23575$+$2536 & 3.97 \\
IRAS~01572$+$5844 & 4.96 \\
IRAS~02547$+$1106 & 3.20 \\
IRAS~03022$+$5409 & 3.94 \\
IRAS~03206$+$6521 & 2.88 \\
IRAS~05131$+$4530 & 2.75 \\
IRAS~05284$+$1945 & 2.82 \\
IRAS~05506$+$2414 & 3.23 \\
IRAS~05552$+$1720 & 2.82 \\
IRAS~06121$+$1221 & 3.65 \\
IRAS~06238$+$0904 & 7.62 \\
0759401$+$152312 & 9.28 \\
IRAS~16030$-$0634 & 4.83 \\
1611445$+$120416 & 3.97 \\
IRAS~16131$-$0216 & 4.58 \\
1644295$+$234759 & 4.19 \\
IRAS~17055$-$0216 & 4.58 \\
IRAS~17193$-$0601 & 4.45 \\
IRAS~17308$+$0822 & 4.58 \\
1744554$+$500239 & 2.91 \\
1749069$+$080610 & 4.38 \\
IRAS~17484$-$1511 & 6.02 \\
IRAS~17531$-$0940 & 5.02 \\
1758333$+$663759 & 3.26 \\
1800071$+$663654 & 4.32 \\
IRAS~18095$+$2704 & 4.16 \\
1812063$+$065113 & 3.33 \\
IRAS~18099$-$1449 & 5.18 \\
IRAS~18100$-$1250 & 4.90 \\
IRAS~18117$-$1625 & 5.09 \\
IRAS~18118$-$1615 & 5.22 \\
IRAS~18123$+$0511 & 3.87 \\
IRAS~18127$-$1516 & 5.18 \\
OH~15.7$+$0.8     & 5.98 \\
1817340$+$100903 & 3.01 \\
IRAS~18156$+$0655 & 5.15 \\
OH~18.8$+$0.4     & 4.42 \\
IRAS~18237$+$2150 & 4.45 \\
IRAS~18236$-$0447 & 4.00 \\
1827092$+$011427 & 5.02 \\
IRAS~18245$-$0552 & 4.80 \\
1829161$+$001822 & 5.09 \\
1829553$+$004939 & 4.96 \\
1830061$+$004233 & 3.81 \\
IRAS~18273$-$0738 & 4.51 \\
1834515$-$081820 & 4.29 \\
1839230$-$055323 & 5.22 \\
IRAS~18501$+$1019 & 3.36 \\
IRAS~18517$+$0037 & 3.71 \\
1854250$+$004958 & 4.35 \\
OH~1854$+$02      & 3.42 \\
IRAS~18587$+$0521 & 5.06 \\
IRAS~18596$+$0605 & 4.96 \\
IRAS~19010$+$0526 & 5.02 \\
IRAS~19017$+$0608 & 4.96 \\
IRAS~19024$+$1923 & 4.70 \\
IRAS~19023$+$0745 & 4.90 \\
IRAS~19027$+$0517 & 5.12 \\
IRAS~19041$+$0952 & 5.12 \\
IRAS~19044$+$0833 & 5.06 \\
IRAS~19047$+$1539 & 4.80 \\
IRAS~19055$+$0225 & 3.39 \\
IRAS~19071$+$0625 & 5.54 \\
1909599$+$043708 & 4.13 \\
IRAS~19079$+$1143 & 5.38 \\
1910544$+$012444 & 4.42 \\
IRAS~19085$+$1038 & 4.74 \\
IRAS~19114$+$0002 & 5.79 \\
IRAS~19117$+$1107 & 3.94 \\
1914408$+$114449 & 4.03 \\
IRAS~19201$+$2101 & 3.14 \\
IRAS~19231$+$3555 & 3.17 \\
IRAS~19283$+$1421 & 3.20 \\
OH~53.6$-$0.2     & 5.02 \\
1932551$+$141337 & 4.29 \\
IRAS~19309$+$2022 & 4.90 \\
IRAS~19315$+$1807 & 4.96 \\
IRAS~19323$+$2103 & 5.09 \\
IRAS~19374$+$1626 & 5.12 \\
IRAS~19414$+$2237 & 5.54 \\
IRAS~19440$+$2251 & 4.90 \\
1949296$+$312716 & 4.58 \\
1952516$+$394326 & 5.22 \\
IRAS~19583$+$1323 & 5.73 \\
2001595$+$324733 & 4.06 \\
IRAS~20010$+$2508 & 5.02 \\
IRAS~20023$+$2855 & 3.68 \\
IRAS~20021$+$2156 & 6.14 \\
IRAS~20020$+$1739 & 6.50 \\
2005300$+$325138 & 6.14 \\
IRAS~20043$+$2653 & 3.87 \\
2010236$+$462739 & 4.70 \\
2012428$+$195922 & 4.99 \\
2013579$+$293354 & 4.80 \\
2015573$+$470534 & 4.61 \\
IRAS~20156$+$2130 & 6.34 \\
IRAS~20181$+$2234 & 6.14 \\
2021328$+$371218 & 3.68 \\
2021388$+$373111 & 5.82 \\
IRAS~20215$+$6243 & 4.00 \\
IRAS~20266$+$3856 & 3.46 \\
IRAS~20305$+$6246 & 4.03 \\
2032541$+$375128 & 4.35 \\
2045540$+$675738 & 4.16 \\
IRAS~20444$+$0540 & 5.31 \\
2048166$+$342724 & 4.48 \\
IRAS~20479$+$5336 & 4.16 \\
IRAS~20523$+$5302 & 4.22 \\
OH~85.4$+$0.1     & 3.94 \\
IRAS~21000$+$8251 & 4.03 \\
IRAS~20549$+$5245 & 4.19 \\
2058537$+$441528 & 4.22 \\
2058555$+$493112 & 4.13 \\
2059141$+$782304 & 4.10 \\
2119074$+$461846 & 4.61 \\
IRAS~21509$+$6234 & 3.97 \\
IRAS~21522$+$6018 & 4.06 \\
2154144$+$565726 & 3.07 \\
IRAS~21554$+$6204 & 3.52 \\
2204124$+$530401 & 3.55 \\
IRAS~22036$+$5306 & 2.43 \\
IRAS~22103$+$5120 & 3.33 \\
2219055$+$613616 & 4.26 \\
2219520$+$633532 & 5.22 \\
2223557$+$505800 & 5.44 \\
2224314$+$434310 & 7.55 \\
2235235$+$751708 & 3.94 \\
IRAS~22345$+$5809 & 2.50 \\
IRAS~22394$+$6930 & 3.81 \\
IRAS~22394$+$5623 & 3.84 \\
2251389$+$515042 & 4.86 \\
IRAS~22517$+$2223 & 3.62 \\
2259184$+$662547 & 3.74 \\
2310320$+$673939 & 2.37 \\
2312291$+$612534 & 2.53 \\
2332448$+$620348 & 6.75 \\
IRAS~23352$+$5834 & 5.06 \\
IRAS~23361$+$6437 & 5.06 \\
IRAS~23489$+$6235 & 3.78 \\
IRAS~23554$+$5612 & 3.68 \\
IRAS~23561$+$6037 & 3.71 \\
\enddata
\end{deluxetable}

\end{document}